\documentclass{jfm}
\usepackage{graphicx}  
\usepackage{natbib}

\usepackage{amsmath}
\usepackage{dcolumn}   
\usepackage{bm}        
\usepackage{amssymb}   
\usepackage{color}
\usepackage[colorlinks=true,citecolor=blue]{hyperref}

\newcommand{\review}[1]{#1}

\newcommand{\reviewsecond}[1]{#1}
\newcommand{\reviewthird}[1]{#1}

\newcommand\bb{\begin{eqnarray}}
\newcommand\ee{\end{eqnarray}}

\newcommand\er{{\bf \hat r}}
\newcommand\bs{\boldsymbol}

\newcommand\bu{{\bf u}}
\newcommand\br{{\bf r}}
\newcommand\hr{{\bf \hat r}}

\newcommand\Ylmf{Y_{\ell}^{m}(\theta,\varphi)}
\newcommand\Ylm{Y_{\ell}^{m}}

\newcommand\alm{a_{\ell,m}}
\newcommand\dalm{\dot a_{\ell,m}}
\newcommand\ddalm{\ddot a_{\ell,m}}

\newcommand\blm{\pi_{\ell,m}}
\newcommand\ulm{u_{\ell,m}}
\newcommand\dulm{\dot u_{\ell,m}}

\newcommand\lm[1]{{#1}_{\ell,m}}
\newcommand\lmf[1]{{#1}_{\ell,m}(\theta,\varphi)}
\newcommand\sumlm{\sum_{\ell=2}^{+\infty} \sum_{m=-\ell}^\ell}
\newcommand\iiOn{\frac{1}{4\pi} \iint \textrm{d}\Omega}

\newcommand\We{\textrm{We}}
\newcommand\zO{\zeta_\Omega}
\newcommand\avg[1]{\langle #1 \rangle}

\begin{document}

\title[Bubble deformation by a turbulent background]{Bubble deformation by a turbulent flow}

\author[St\'ephane Perrard \textit{et al.}]
{St\'ephane Perrard$^{1,2}$, Ali\'enor Rivi\`ere$^{1,2}$, Wouter Mostert$^{1}$, Luc Deike$^{1,3}$}

\affiliation{$^1$Department of Mechanical and Aerospace Engineering, Princeton University\\[\affilskip]
$^2$LPENS, D\'epartement de Physique, Ecole Normale Sup\'erieure, PSL University, 75005 Paris, France \\[\affilskip]
$^3$ Princeton Environmental institute, Princeton University}

\date{22/05/2020; revised 12/07/2020, 08/01/2021, 08/03/2021; accepted 13/04/2021}

\maketitle

\begin{abstract}
\review{We investigate the modes of deformation of an initially spherical bubble immersed in a homogeneous and isotropic turbulent background flow. \reviewthird{We perform direct numerical simulations of the two-phase incompressible Navier-Stokes equations, considering a low-density bubble in the high density turbulent flow at various Weber number (the ratio of turbulent and surface tension forces) using the air-water density ratio}. We discuss a theoretical framework for the bubble deformation in a turbulent flow using a spherical harmonic decomposition. We propose, for each mode of bubble deformation, a forcing term given by the statistics of velocity and pressure fluctuations, evaluated on a sphere of the same radius. This approach formally relates the bubble deformation and the background turbulent velocity fluctuations, in the limit of small deformations. The growth of the total surface deformation and of each individual mode is computed from the direct numerical simulations using an appropriate Voronoi decomposition of the bubble surface. We show that two successive temporal regimes occur: the first regime corresponds to deformations driven only by inertial forces, with the interface deformation growing linearly in time, in agreement with the model predictions, whereas the second regime results from a balance between inertial forces and surface tension. The transition time between the two regimes is given by the period of the first Rayleigh mode of bubble oscillation. We discuss how our approach can be used to relate the bubble lifetime to the turbulence statistics and eventually show that at high Weber number, bubble lifetime can be deduced from the statistics of turbulent fluctuations at the bubble scale.}
\end{abstract}

\section{Introduction}

The interaction of turbulent flow with a free surface occurs in numerous physical systems, from ocean waves forced by the turbulent wind~\citep{Phillips_1957,Perrard_JFM_2019}, to river surface patterns driven by underwater turbulence~\citep{Peregrine_1976,Brocchini_2001_1}, to drops and bubbles in turbulent flow~\citep{Eaton2010,Elghobashi2019,Lohse2020}.

Deformation and break-up of bubbles and droplets in a turbulent flow control exchanges of heat, mass and momentum in numerous natural and engineering processes, from bubble-mediated gas exchange at the ocean-atmosphere surface~\citep{Deike2018,Reichl2020}, chemical reactors~\citep{Risso2018}, to the fall of rain drops~\citep{Villermaux2009}, to the evaporation of sea spray in a turbulent boundary layer~\citep{Veron2015}, to dynamics of water droplets in clouds~\citep{Eaton2010} to industrial liquid atomization and fragmentation~\citep{Eggers2008}. 

The dynamics of a single bubble evolving freely in low-viscosity liquid at rest have been extensively studied, with analytical and experimental results describing bubble oscillations \citep{Prosperetti_1980,Miller1968}, rise velocity \citep{Moore1965,Maxworthy1996} and path instability \citep{Magnaudet_2000,Ern2012}. The rise dynamics are altered once bubbles are close enough to each other and can interact through their wakes \citep{Harper1970,Yuan1994}, which at high void fraction can interact through their wakes, a process that leads to collective effects and bubble induced turbulence dynamic \citep{Lance1991,Risso2018}.

The deformation dynamics and potential break-up of a bubble depend primarily on the ability of the surrounding fluid to deform the bubble against surface tension forces. This defines the Weber number, comparing inertial forces generated by the turbulent carrier flow and the capillary cohesive forces. Considering the velocity fluctuation at the bubble diameter scale $d_0$, formalized by the longitudinal velocity increment $\delta u({\bf d_0}) = u_L(\br,t) - u_L(\br+{\bf d_0},t)$, \review{where $u_L$ is the velocity component along the direction of $\bf d_0$}, the turbulent Weber number is defined as $\We = \rho_\ell \langle \delta u(d_0)^2 \rangle d_0/\gamma$~\citep{Hinze1955,Risso_1998} with $\rho_\ell$ the density of water, $\gamma$ the air-water surface tension and $\langle \rangle$ the average over the flow configurations. In a homogeneous and isotropic turbulent flow, the velocity fluctuations at the bubble scale $\delta u(d_0)^2$ can be related to the mean dissipation rate of energy $\epsilon$ using \citet{K41} theory, yielding $\langle \delta u(d_0)^2 \rangle = C (\epsilon d_0)^{2/3}$ for $d_0$ in the inertial range. Experimental studies have observed $C \in [2,2.2]$ depending on Reynolds number~\citep{Variano_2008,Pope_book}. We chose $C=2$ for consistency with~\citet{Risso_1998}, and the Weber number eventually writes:
\bb
\We = \frac{2\rho_\ell \epsilon^{2/3}d_0^{5/3}}{\gamma},
\label{We}
\ee
for a bubble of diameter $d_0$ immersed in a homogeneous and isotropic turbulent flow. \review{Experimental studies of bubble dynamics in turbulence have identified a critical Weber number $\We_c$ of order unity, above which the occurence of break-up becomes statistically dominant. The value of $\We_c$ reported from laboratory experiments varies among authors in the range [1,5], corresponding to variation in experimental conditions, which introduces other flow parameters such as large scale shear or spatial variations in the dissipation rate ~\citep{Lasheras1999,Andersson2006,Ravelet_2011,vejravzka2018experiments}. Note also that the critical Weber number is defined statistically, and can be influenced by the temporal and spatial windows of observation of the bubbles. In this article, we consider $\We_c = 3$, in accordance with our numerical dataset, obtained from ensemble of simulations.} Two mechanisms driving the deformation and break-up have been discussed~\citep{Lasheras1999,Andersson2006,Ravelet_2011,vejravzka2018experiments}, either the direct strong action of an eddy at the scale of the bubble leading to large deformation and break-up, or a resonance mechanism between deformation caused by weaker eddies and oscillations of the bubble~\citep{Risso_1998}. Experimental studies have identified an oscillatory response of bubbles in turbulence associated to the second eigenmode \citep{Risso_1998,Ravelet_2011}. \reviewthird{Such oscillatory response of millimetric bubbles, associated with surface tension forces, is characteristic of the large deformation observed prior to break-up, and corresponds to a much lower frequency than the acoustic mode of deformation, characterized by the Minnaert frequency~\citep{Minnaert_1933,Deane_2002}.}




In such configurations, the gas-liquid interface is surrounded by a stochastic turbulent flow, characterized by perturbations of various strengths at various scales. \reviewsecond{In this paper, we discuss a theoretical framework of bubble oscillations to describe the temporal evolution of the bubble interface deformation while immersed in a turbulent background flow. To do so, we consider linearized deformations and we neglect any feedback of the interface on the statistics of the turbulent background flow. Our framework has two main inspirations. On one hand, it extends the spherical harmonic decomposition of bubble deformation~\citep{Miller1968,Prosperetti_1980} to the presence of external forcing. On the other hand, it is analogous to a liquid-gas interface interpreted as a collection of harmonic oscillators which are forced by a turbulent background flow \citep{Risso_1998,lalanne2019model}. This is similar to the case of wind wave generation by pressure fluctuations in a turbulent boundary layer, as proposed by \citet{Perrard_JFM_2019} and built on earlier work by~\citet{Phillips_1957}}. However, as we will see, the solutions describing the growth of the bubble interface deformation lie in a different dynamical regime, in which a large separation of time scales between stochastic growth and saturation is not fulfilled, in part because the turbulent forcing occurs in the dense liquid, imposing a rapid response from the gas phase, contrary to forcing in the turbulent air of much weaker inertia inducing deformation at the surface of the dense liquid phase. 

\reviewsecond{We aim to estimate the turbulent excitation in terms of pressure and velocity statistics evaluated on the bubble surface. The utility of the present theoretical framework hence lies on the ability to test numerically the relationship between surface deformation and turbulent background flow statistics.} 

We discuss an equation linking the growth of the bubble deformation decomposed in spherical harmonics to the turbulence statistics. Doing so, we identify different regimes. At short time, for $t\ll t_2,t_c$, where $t_2$ is the capillary time describing the oscillatory response of the bubble, and $t_c$ is the eddy turn-over time at the scale of the bubble, the bubble response is independent of surface tension (Weber number) and the total bubble deformation $\zeta_{\Omega}$ follows a linear scaling $\zeta_{\Omega} \sim t$ . The amplitudes of the spherical harmonics coefficients also follow a linear growth in time, while the response is dominated by the modes 2 of oscillation. At intermediate time scale $t_2<t<t_c$, bubble deformations are described by a sub-linear regime, where the transition time is a function of surface tension (Weber number). The model is evaluated by direct numerical simulations of bubble deformation and break-up in a turbulent flow, resolving the full two-phase air-water Navier-Stokes equations. We compute the bubble deformation, as well as the evolution of the various dynamical modes of deformation. We observe very good agreement between the simulations and the model and observe the different regimes discussed theoretically. Finally at long time, the bubble deformation either saturates or the bubble breaks, both regimes not being described by our linear model but captured in the simulations. We eventually link the bubble deformation growth to the statistical properties of the velocity increments. Doing so, we provide an estimate of the bubble lifetime statistics, which is parametrized by the Reynolds number and the Weber number.

The paper is organized as follows, in \S 2, we present the theory, while in \S 3 we present the simulations and comparison with the theoretical prediction. Conclusions and discussions are presented in \S 4.

\section{Bubble deformation in a turbulent flow\label{theory}}
	\begin{figure}
\centering
\includegraphics[width = 0.8 \linewidth]{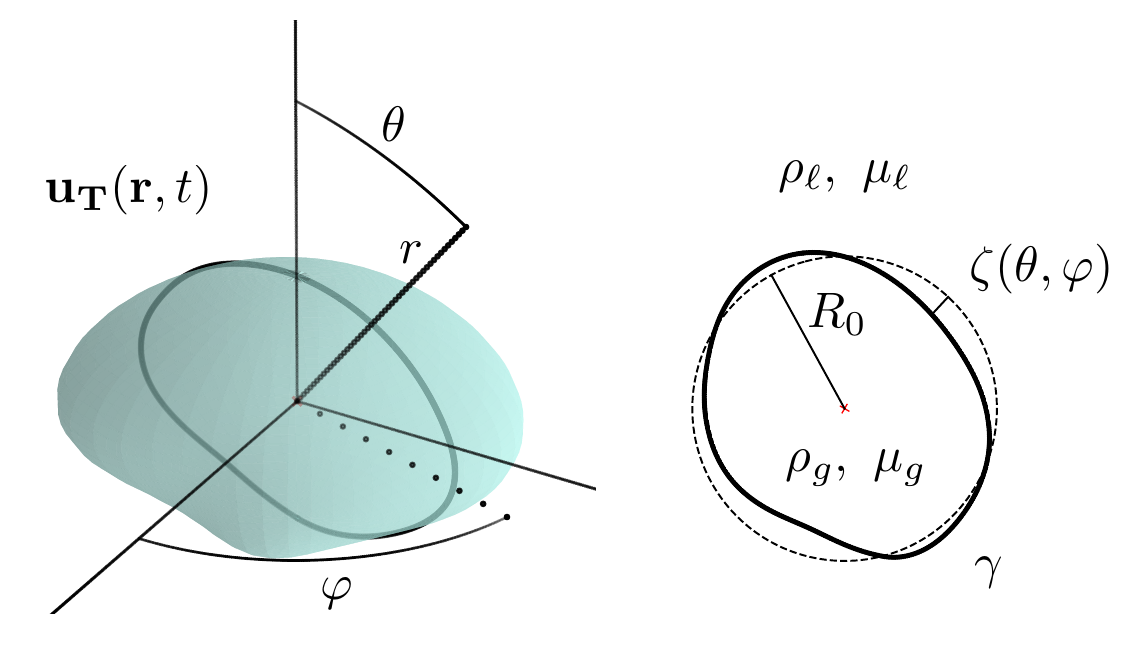}
\caption{Bubble deformation in a turbulent flow : sketch \& definitions.}
\label{figsketch}
\end{figure}
	\subsection{Bubble oscillation in a fluid at rest}
\reviewsecond{We consider an \reviewthird{incompressible}, initially spherical gas bubble of radius $R_0=d_0/2$, density $\rho_g$ and viscosity $\mu_g$, immersed in a liquid at rest of liquid density $\rho_\ell$, viscosity $\mu_l$, and surface tension between the gas and the liquid phase $\gamma$. The origin of space is set at the centre of mass of the bubble. \reviewthird{The chosen frame of reference is then non inertial. Note that we do not study here the motion of the centre of mass and aim at description the bubble deformation.} Focusing on the early stage, we limit our description to a bubble surface that can be parametrized in spherical coordinates $(r,\theta,\varphi)$ by the single-valued distance ${\bf R} = (R_0(t) + \zeta(\theta,\varphi,t)) {\bf \hat r} $ to the center (Figure 1). We aim for a phenomenological set of equations for the modes of deformation of the bubble in the presence of a turbulent background flow, inspired by the forced oscillator model from \citet{Risso_1998,lalanne2019model}. 

	To decompose the bubble surface into a basis adapted to the spherical geometry, we introduce the spherical harmonic functions $\lmf{H}$ :
\begin{equation}
\lmf{H} = \sqrt{\frac{(2\ell+1)(\ell-m)!}{(\ell+m)!}} P^m_\ell(\cos{\theta}) e^{i m\varphi} 
\end{equation}
with $\ell \in \mathbf{N}$ and $m \in \left[-\ell, \ell \right]$, and $P^m_\ell$ is the associated Legendre polynomial. The real form $\lmf{Y}$ of the spherical harmonics writes:
\begin{equation}
    \Ylm = \left\{
    \begin{array}{ll}
        \sqrt{2}(-1)^m \mathbf{Im}(H_{\ell}^{-m}) &\mbox{if } m<0 \\
        H_\ell^0 & \mbox{if } m=0 \\
        \sqrt{2}(-1)^m \mathbf{Re}(H_{\ell}^{m})    &\mbox{if } m>0. \\
    \end{array}
\right.
\end{equation}
The real spherical harmonics $\Ylm$ verifies $\forall \ell \in \mathbb{N}, \forall m \in [-\ell,\ell], \nabla^2 r^{-\ell} \Ylm = 0$ for $r>0$. In this basis, the expression of the bubble surface $R(\theta,\varphi,t)$ writes:
	\bb
	R(\theta,\varphi,t) &=& R_0(t) + \sumlm \alm(t) \Ylmf  	\label{sphharm} \\
	\label{sphharm2}
	\ee
where $R_0(t)$ is the mean bubble radius. In the following, we neglect the radius variation associated with gas compressibility and we assume that $R_0$ is independent of time. In the absence of any background flow, at the linear order in the deformation amplitude $\zeta/R_0$ and small viscous dissipation in the bubble boundary layer, the coefficients $\alm$ are solutions of a set of oscillator equations of the form~\cite{Prosperetti_1980}:
\bb
\ddalm + 2 \beta_\ell \dalm + (\ell-1)(\ell+1)(\ell+2) \omega_0^2 \alm = 0.
\label{alm_eq0}
\ee
where $\cdot$ denotes the time derivative, $\omega_0 = \sqrt{\gamma /(R_0^3\rho_\ell)}$ is the typical angular frequency associated with bubble oscillation and $\beta_\ell$ is the damping coefficient of mode $\ell$. The expression of $\beta_2$ is summarized in~\citet{lalanne2019model} for low viscosity, and the general formula can be found in~\citet{Miller1968}. We introduce $\omega_\ell = \omega_0 \sqrt{(\ell-1)(\ell+1)(\ell+2)}$ the angular frequency associated to mode $\ell$ and the associated characteristic time $t_\ell = 1/\omega_\ell$ of mode $\ell$. We have in particular $\omega_2 = 2 \sqrt{3} \omega_0$.

\subsection{Bubble oscillation in a turbulent background flow}

	We consider that the bubble is immersed in a liquid animated by turbulent motion, with a velocity $\bs u(\bs r,t)$ and a pressure field $p(\bs r,t)$. In the general case, modelling the action of the turbulent background flow on a bubble is particularly complex. A two-way coupling between the velocity fluctuations near the bubble interface and the interface deformation may take place. The turbulent flow deforms the bubble under the action of normal and tangential stresses, and the turbulent flow statistics can be modified in the immediate vicinity of the bubble interface. In order to construct a phenomenological model for the modes of deformation of the bubble, we follow the approach of \citet{Risso_1998,lalanne2019model}, which considers a dynamical equation for the second mode of deformation $a_2$, assumed to be the main mode excited by the flow. The amplitude $a_2(t)$ then represents the oscillation of the bubble between an oblate and a prolate shape.
	
	We introduce two adaptations of the formulation of ~\citep{Risso_1998} and \citep{lalanne2019model}. i) We consider that the turbulent flow can excite other modes $(\ell,m)$ of deformation. The contribution of these higher order modes $(\ell > 2)$ could be non-negligible for high Weber numbers ($We \gg We_c$). ii) We aim to estimate the turbulent excitation in terms of pressure and velocity statistics evaluated on a sphere of the same radius, in the bulk of the turbulent flow. We perform a linear expansion in the amplitude of the modes of deformation, and we neglect the effect of the interface on the turbulent statistics in its immediate vicinity. The decoupling approach was first motivated by the high Weber number case, in which capillary forces are small compared to inertial forces. For lower Weber number ($We \approx We_c$), our approach is expected to still hold at short time compared to the response time $1/\omega_\ell$. We also consider large dissipative time compared to the bubble period of oscillation $(\omega_\ell/\beta_\ell \gg 1)$, such that we neglect the viscous dissipation. For all these reasons, the following approach is rather phenomenological and by no means exact. The validity of our approach is further discussed by comparison with direct numerical simulations of a single bubble of initial spherical shape deformed by a homogeneous and isotropic turbulent flow.

	At the linear order of deformation ($\alm/R_0 \ll 1$), we assume that the bubble surface can still be described by a set of oscillatory equations, adding a forcing term that represents the turbulent background flow
\bb
\ddalm + \omega_\ell^2 \alm = \lm{F}
\label{alm_eqF}
\ee
where $\lm{F}$ is a function of $\bs u(\bs R,t)$, $d\bs u/dt(\bs R,t)$ and $p(\bs R,t)$, respectively the velocity field, the acceleration field and the pressure field evaluated at the surface of the bubble. We neglect the shear contribution and, looking at the early time dynamics, we neglect viscous dissipation. To specify the function $\lm{F}$, we introduce the spherical harmonic decomposition of $u_r(R_0 \hr,t) = {\bf u} \cdot {\hr}$ and $p(R_0 \hr,t)$ evaluated on a sphere of radius $R_0$:
\bb
u_r(R_0 \hr,t) &=& \sumlm \lm{u}(t) \Ylmf, \\
p(R_0 \hr,t) &=& \rho_\ell \sumlm \lm{\pi}(t) \Ylmf,
\ee
where $\hr$ is the unit vector along $\bs r$. We then approximate the fields $\bs u(\bs R,t)$, $d\bs u/dt(\bs R,t)$ and $p(\bs R,t)$ by their value on an undeformed bubble, whose interface position is $R_0 \hr$. Doing so, we neglect the contribution of cross terms of the type $\alm \ulm$ and terms of $\alm$, $\ulm$, $\blm$ of quadratic and higher order. In this limit, the coefficients of the spherical harmonic decomposition of $\dot u_r(R_0 \hr,t)$ become $\lm{\dot u}$ as $R_0 \hr$ is independent of time. We also neglect the coupling between two different spherical harmonic modes $(\ell,m)$ and $(\ell',m')$ which arises at least at the order $\alm u_{(\ell',m')}$ or $\alm a_{(\ell',m')}$. For $\alm/R_0 \ll 1$, we then assume that $\lm{F}$ is a linear function of $\ulm, \lm{\dot u}$ and $\lm{\pi}$, and we postulate the following form:
\bb
\ddalm + \omega_\ell^2 \alm = \lm{\dot u} + \frac{C(\ell)}{R_0} \lm{\pi},
\label{alm_eq1}
\ee
where $C$ is a function of $\ell$. The kinematic boundary conditions at $t=0$ give $R(\theta,\varphi,0) = R_0$ and $\dot R(\theta,\varphi,0) = u_r(R_0,\theta,\varphi,0)$, which gives the initial conditions for $\alm$,
\bb
\alm(0) = 0;  \text{ and }
\dalm(0) &=& \ulm(0).
\ee}

\subsection{Link with surface deformation}

We seek for an estimate of the average global surface deformation. We introduce the root mean squared $\zO$ of the bubble interface from an integration over all solid angles:
    \begin{equation}
    \zO^2 = \iiOn \zeta^2.
    \label{zetarms}
    \end{equation}
According to the orthogonality of spherical harmonics $\iiOn \Ylm Y_{\ell'}^{m'} = \delta_{\ell,\ell'} \delta_{m,m'}$, $\zO$ can be expressed as
\begin{equation}
    \zO^2(t) = \sumlm \alm(t)^2.
\end{equation}
The surface deformation is then related to the solutions of Eq.~\ref{alm_eq1}. $\alm$ being stochastic variables, we look for its temporal behaviour as well as temporal behavior of its ensemble averaged value $\langle \alm^2 \rangle$.
	
\subsection{Controlling time scales}
\begin{figure}
\centering
\includegraphics[width = 0.6 \linewidth]{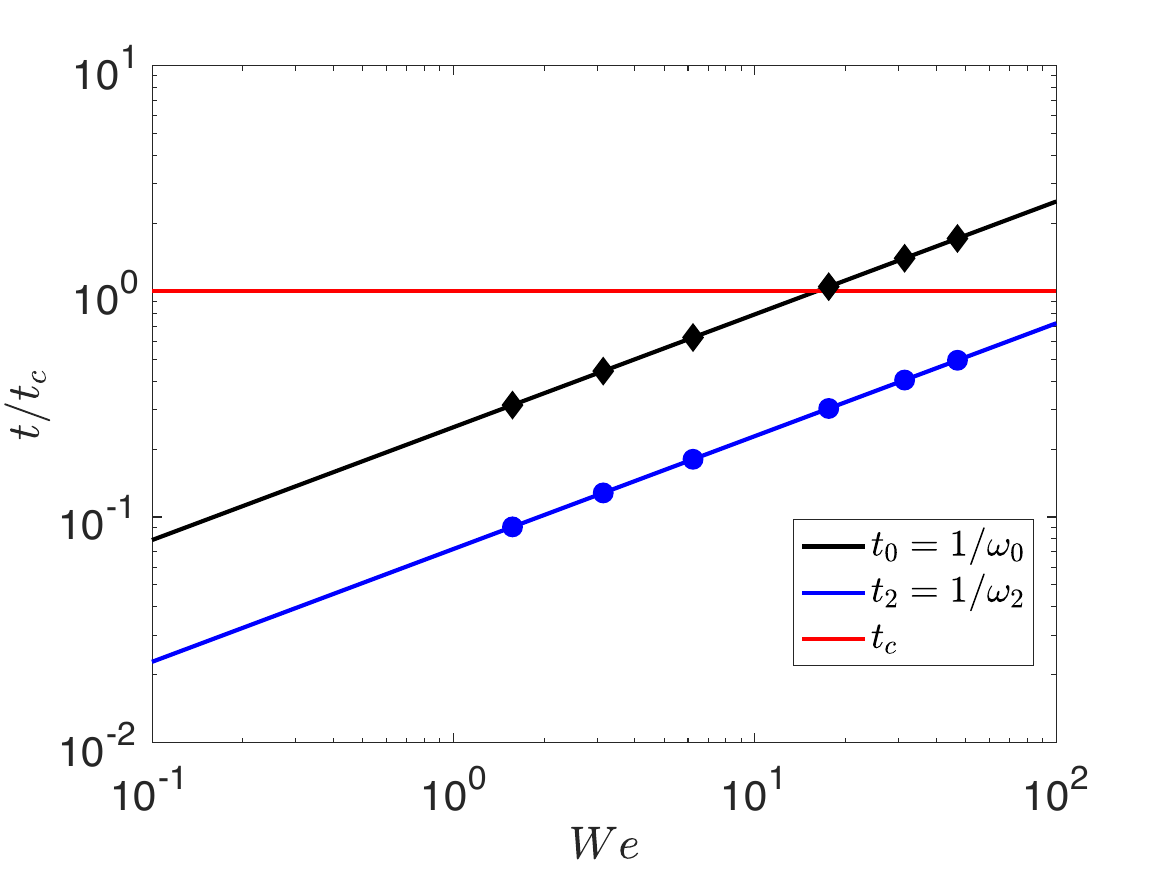}
\caption{Ratio of characteristic bubble oscillation time scales with the turbulence time scale of the problem as a function of the Weber number. The turbulence time scale is the eddy turn over time at the bubble scale, $t_c =\epsilon^{-1/3} d_0^{2/3}$, the capillary oscillation time is $t_0=1/\omega_0 =1/\sqrt{\gamma /(R_0^3\rho_\ell)}$ and the time associated to mode 2 is $t_2=1/\omega_2 = 1/(2 \sqrt{3} \omega_0)$. Symbols correspond to the DNS cases in section 3.}
\label{figtime}
\end{figure}
The solutions of Eq.~\ref{alm_eq1} depend on the various time scales that are inherently associated to a noise-excited oscillator. For a bubble evolving in a turbulent flow, previous studies have clearly identified one controlling time scale, the natural bubble oscillation response occurring at a capillary time scale~\citep{Risso_1998,Ravelet_2011}. Another relevant time scale lies in the turbulence correlation time of the fluctuations at the bubble scale also discussed by several authors~\citep{Risso_1998,Lasheras1999,Ravelet_2011}. Depending on the Weber number, another time scale is eventually given either by the bubble lifetime or the saturation time.
	
\review{The modes $\ell=2$ carry most of the surface energy as they correspond to the largest available spatial scale. Therefore, the typical characteristic amplitude and time scales of $\zO$ are set by the modes $\ell=2$ of oscillation.} The first relevant time scale is the oscillator reduced period $t_{2}=1/\omega_2$ of the second mode $\ell=2$ of oscillation, which has been proven to dominate the deformation~\citep{Risso_1998}. On a time significantly shorter than the reduced period $\omega_2 t < 1$, the response will be purely inertial and independent of surface tension~\citep{Risso_1998}. 
	
The second characteristic time scale arises from the turbulence temporal correlations $\lm{\tau}$ and $\lm{\tilde\tau}$ of respectively the coefficients $\ulm$ and $\lm{\pi}$ of the spherical decomposition of the velocity and pressure fluctuations. They can be defined from the autocorrelation functions of velocity and pressure respectively :
\bb
\lm{\tau} &=& \frac{1}{\avg{\ulm^2}}\int_{-\infty}^{+\infty} \textrm{d}s ~ \langle \ulm(t) \cdot \ulm(t+s) \rangle. \\
\lm{\tilde\tau} &=& \frac{1}{\avg{\lm{\pi}^2}} \int_{-\infty}^{+\infty} \textrm{d}s ~ \langle \lm{\pi}(t) \cdot \lm{\pi}(t+s) \rangle.
\ee
\reviewthird{The estimate of $\lm{\tau}$ presents some difficulties, as there is no clear consensus on the time scales in turbulence. The velocity fluctuations we consider here are neither associated with the Eulerian frame of reference nor a local Lagrangian framework. To provide an estimate of the time scale $\tau_{2,m}$ of the dominant mode of deformation, we consider the time scales $t_E$ and $t_L$ respectively of the Eulerian velocity increments and the Lagrangian velocity increments at the bubble scale. The ratio $t_E/t_L$ scales as $(Re_\lambda)^{1/2}$ from the random sweeping hypothesis~\citep{Tennekes_1975}, such that $t_E$ and $t_L$ can respectively be considered as lower and upper bounds of the correlation time $\tau_{2,m}$. The frozen Taylor hypothesis will henceforth be considered as valid for $t<t_E$, and we estimate $\tilde\tau_{2,m}$ from the eddy turn over time at the bubble scale $t_c = \epsilon^{-1/3} d_0^{2/3}$. The correlation times $\tau_{\ell,m}$ and $\tilde\tau_{\ell,m}$ of the higher order modes ($\ell > 2$) are expected to scale as $\ell^{-2/3}$ whereas the capillary time 1$/\omega_\ell$ scales as $\ell^{-3/2}$. The criterion $\omega_2 \tau_2 < 1$ therefore implies $\omega_\ell \tau_\ell < 1$ for higher values of $\ell$.}

Figure 2 shows the ratio of the typical capillary time scales ($t_0$ and $t_2$) with the eddy turn-over time scale $t_c$ as a function of Weber number. Below the critical value $\We<\We_c$ with $\We_c=3$, we have $t_2\ll t_c$, and this separation persists at moderate We while being reduced. The crossover between these two time scales, $t_2=t_c$ occurs at higher We number, exactly $\We = 192$ (and $\omega_2=\sqrt{192}/(t_c\sqrt{We})$). For reference, we provide typical time scales in table \ref{Table1}, corresponding to experimental conditions of bubbles evolving in turbulent flow created from underwater turbulent jets~\citep{Ruth_PNAS_2019,Risso_1998,Lasheras1999,vejravzka2018experiments}. Note that these experimental systems can present large variations in turbulence levels within the flow, and significant mean shear. A complete discussion on the challenges of experimental systems to study bubble deformation and break-up in homogeneous and isotropic turbulence can be found in \citet{Ni2019}. The long time evolution of bubble deformation depends eventually on the bubble stability. The stability being dependent on the deformation, we postpone the discussion on either the saturation or the break-up time to the end of the direct numerical simulations section \S 3. The following sections derive asymptotic solutions associated to each temporal regimes.

\reviewthird{We note that the compressible mode of deformation for a bubble or diameter $2R$ in a infinite domain of water oscillates at the Minnaert frequency $f_M = 1/(2\pi R) (3 \Gamma p_0/\rho_\ell)^{1/2} $ \citep{Minnaert_1933}, where $\Gamma$ is the polytropic coefficient and $p_0$ the bubble pressure. When considering millimetric air bubble in water, the Minnaert frequency $f_M$ ranges from $[0.5,10]$~kHz for sizes in the range $[10,0.5]$~mm~\citep{Deane_2002}. The acoustic frequency $f_M$ can be compared to the Lamb oscillation frequency~$1/t_2=\omega_2/(2 \pi)$~\citep{Lamb_1995} which is driven by surface tension forces and characteristic of the large deformation and break-up dynamics. For an air bubble in water of millimetric size, this second mode of oscillation $1/t_2$ ranges between [0.025,1]~kHz, so that the ratio $f_M t_2 > 10$ and air compressibility can be neglected when analyzing large deformations leading to break-up.}



\begin{table}
\centering
\begin{tabular}{p{4cm} p{1.4cm} p{1.4cm} p{1.4cm} p{1.4cm} p{1.4cm} }
Reference & $\epsilon$ (m$^2$s$^{-3}$) & $L$ (mm) & $d$ (mm) & $t_2$ (ms) & $t_c$ (ms) \\
\hline
\citet{Ruth_PNAS_2019} &  0.5-1 & 13 & 2-6 & 1-6 & 20-40\\
\citet{Risso_1998} &  0.1-2 & 21 & 4-16 & 2-40 & 50-140 \\
\citet{Ravelet_2011} &  1 & 10 & 9 & 10 & 50 \\
\citet{Lasheras1999} &  3-100 & 48 - 115 & 3 & 2 & 5-35 \\
\citet{vejravzka2018experiments} &  0.1-10& 4-6  & 2-6 & 1-6 & 15-140 \\
\citet{Ni2019} &  0.16 &45-60 & 2-6 & 1-6 & 15-30 \\
\hline
\end{tabular}
\caption{Estimated experimental time scales and length scales for typical laboratory experiments of bubble deformation and break-up in turbulence. All experiments are made with air bubbles in water. Dissipation rate $\epsilon$ and bubble diameter $d$ were directly provided by all authors. An estimate of the integral length scale $L$ were provided by most of the authors, we infer the values for \citet{Lasheras1999} using the usual statistical properties of a turbulent jet~\citep{Hussein_1994}. The oscillation time is given by $t_2=\frac{1}{2 \sqrt{3}\sqrt{\gamma /(R_0^3\rho_\ell)}}$ and the eddy turn over time at the bubble scale is given by $t_c = \epsilon^{-1/3} d_0^{2/3}$. \label{Table1}}
\end{table}

\subsection{Early time solution: linear regime}

Starting from a spherical shape at $t=0$, the first instants verify $\omega_\ell t \ll 1$ and $t \ll t_c$. In this limit, we can consider $\ulm$ and $\lm{\pi}$ as independent of time, and neglect the surface tension term. From eq.~\ref{alm_eq1} and its initial conditions, we get by double integration over time:
\bb
\alm = \ulm t + \frac{C(\ell)}{2 R_0} \lm{\pi} t^2.
\label{alm_lin}
\ee
The relative influence of each term will depend on the characteristic time scale $2 R_0 \ulm/\lm{\pi}$, of order $t_c$ as it also corresponds to a characteristic time scale of the turbulence at the bubble scale $R_0$. For $t \ll t_c$, the pressure contribution is then negligible and the dynamics of each mode $\alm$ reduce to a simple advection by the background flow. Performing an ensemble average operation yields:
\bb
\avg{\alm^2} &=& \avg{\ulm^2} t^2,
\label{alm_lin_avg}
\ee
At short time ($\omega_2 t \ll 1$, $t \ll t_c$), the evolution of $\alm$ is then expected to be independent of surface tension and pressure fluctuations, and this first linear growth regime should hold for all Weber numbers. Note that this limit is similar to the frozen turbulence regime identified for bubble pinch-off in turbulence, in which most of the pinching dynamics occur on a fraction of the eddy turn over time at the scale of the neck~\citep{Ruth_PNAS_2019}. In both cases, the influence of the turbulent background flow reduces to the set of the initial conditions.

Later on, the linear growth regime is limited in time by either the bubble natural oscillations at $1/\omega_2$ or by the correlation time of the turbulent flow $t_c$,  depending on the Weber number as shown in fig.~\ref{figtime}, or by viscous effects which are not investigated here. For $\We \gg 1$, when $t_c < t_2$, the linear regime occurs until $t \approx R_0/\ulm' \approx t_c$, which corresponds to $\zO/R_0 \approx 1$, a deformation amplitude sufficient to trigger bubble breaking. $t_c$ then coincides with the bubble lifetime, and no later time evolution is expected. 
		
\subsection{Forced oscillator regime}

	We consider $\We < 100$, in which the linear growth is first limited by surface tension ($\omega_2 t_c >1$). For $t < t_c$, we still consider that $\ulm$ and $\lm{\pi}$ are independent of time. We have:
\bb
\ddalm + \omega_\ell^2 \alm &=& \dulm +  \frac{C(\ell)}{R_0} \lm{\pi} \\
\dalm(0) &=& \ulm
\ee
which admits a solution of the form:
\bb
\alm =  \frac{\ulm}{\omega_\ell} \sin(\omega_\ell t) + \frac{C(\ell) \lm{\pi}}{R_0 \omega_\ell^2} (1 - \cos(\omega_\ell t))
\ee
The evolution of $\langle \alm^2 \rangle$ at short time is given by :
\bb
\langle \alm^2 \rangle =\langle \ulm^2 \rangle t^2 + \frac{C(\ell) \langle \blm \ulm \rangle}{R_0} t^3  - \frac{1}{3} \omega_\ell^2 \langle \ulm^2 \rangle t^4 +  \frac{C(\ell)^2 \langle \blm^2 \rangle}{R_0^2} \frac{t^4}{4} + O(t^5),
\ee
However, since the terms in $\avg{\blm \ulm}/R_0$ and $\avg{\blm^2}/R_0^2$ are respectively of order $\langle \ulm^2 \rangle /t_c$ and $\langle \ulm^2 \rangle /t_c^2$, their contribution can be neglected for $t \ll t_c$. The first relevant non linear term in the range $1/\omega_\ell < t < t_c$ is then the term $-1/3 \langle \ulm^2 \rangle t^4$, and the equation for $\alm$ becomes:
\bb
\langle \alm^2 \rangle =   \langle \ulm^2 \rangle t^2 - \frac{1}{3} \omega_\ell^2 \langle \ulm^2 \rangle t^4 + O(t^5),
\label{complex_fit}
\ee
which corresponds to a saturation of the linear growth by capillary forces on a time scale $1/\omega_\ell$. 

\section{Direct Numerical simulation of bubble deformation in a turbulent flow\label{DNS}}

\subsection{Numerical methods: the Basilisk flow solver}

We perform direct numerical simulations of the three-dimensional, incompressible Navier-Stokes equations, either with a single phase (the turbulence precursor simulation) or with two-phases (air bubble and turbulent water) with surface tension, using the free software Basilisk \footnote{http://basilisk.fr/} \citep{Popinet2009,Popinet2018}. We use a spatial adaptive octree grid allowing to save computational time while resolving the different length scales of the problem and a momentum conserving scheme. The interface is reconstructed by a sharp geometric Volume of Fluid (VOF) method~\citep{Popinet2009,Popinet2018}. The solver has been extensively described in recent publications~\citep{Popinet2015,Popinet2018,VanHooft2018,Fuster2018,Mostert2020}, and its accuracy has been largely validated on complex multiphase flow, including bubble dynamics~\citep{cano2016paths}, bubble bursting~\citep{Deike2018bursting,Lai2018,Berny2020}, and wave breaking~\citep{Deike2016,Mostert2020,Mostert2020b}. We do not consider the effect of gravity in this work. The turbulent two-phase simulations of bubble deformation in turbulence are presented below.

\subsection{Preparation: creation of the turbulence and insertion of the bubble}
	
	The turbulent flow is generated by adding to the Navier-Stokes equation a volumetric forcing term locally proportional to the velocity field $f \bu$, following the approach of~\citet{Meneveau}, previously implemented and provided as an example in the Basilisk library\footnote{http://basilisk.fr/src/examples/isotropic.c}. ~\citet{Meneveau} showed that such forcing in every point of the real space leads to a well-characterized homogeneous and isotropic turbulent flow with properties similar to those obtained with a spectral code and forcing. Such an approach has also been used by~\cite{Loisy2017} to study rising bubbles in a turbulent flow. 

We consider a cubic box of size L with periodic boundary conditions on each side. Adaptive mesh refinement is used on the velocity field, and the maximum level of refinement $N$ allows comparison with a fixed grid resolution having $2^N$ grid points in each direction. The turbulent flow is generated for increasing resolutions with $N$ going from 6 to 8, corresponding to equivalent $64^3$ to $256^3$ grid size on a fixed grid. This resolution is modest but will be increased around the interface when we inject the bubble in the flow, and is chosen to keep the computational cost reasonable for each simulation as we aim to perform a large number of simulations to access statistical properties of the bubble deformation dynamics.

\begin{figure}
\centering
\includegraphics[width=0.95\columnwidth]{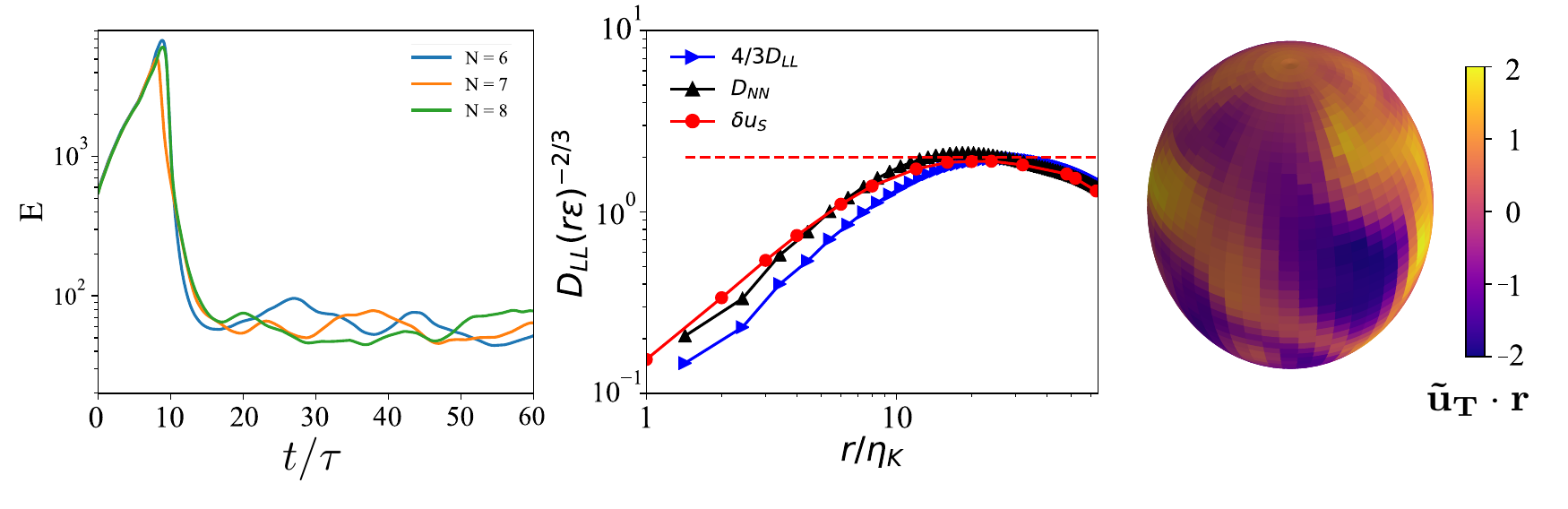}
\caption{(a) Kinetic energy as a function of time in the precursor simulation, for increasing numerical resolution, respectively level N=6, 7 and 8. A statistically stationary state is reached after 15$\tau$ and numerical convergence is reached for level N=7, with a corresponding turbulent Reynolds number Re${_{\lambda}}$=38. A bubble will be immersed in this statistically stationary flow, using different initial times as initial conditions. (b) Second order structure function $D_{LL}$ and $D_{NN}$ in the longitudinal and transverse directions respectively, compensated by the homogeneous and isotropic turbulence scaling $(r \epsilon)^{-2/3}$ and $D_{LL} = 3/4 D_{NN}$. Kolmogorov theory is superimposed in the red dashed line. (c) Visualization of a snapshot of the instantaneous radial velocity component of the turbulent flow evaluated on a sphere of radius $R/\eta_k=32$ that a bubble of corresponding size will face.}
\label{figturb}
\end{figure}

Figure \ref{figturb}a shows the time evolution of the kinetic energy in the precursor simulation, $K = \frac{1}{V}\iiint \frac{1}{2}\rho_\ell u(\vec{x}, t)^2 \mathrm{d}V$. After a short transient, injection and dissipation of energy eventually balances on average and we obtain a statistically stationary, homogeneous and isotropic turbulence. The statistically stationary state is reached after approximately 15 eddy turn over times at the integral scale $\tau = u'^2 / \epsilon$, where $\epsilon$ is the asymptotic dissipation rate and $u'$ the asymptotic root mean squared velocity. The associated Taylor Reynolds number is $Re_{{\lambda}} = \frac{2 K}{3 \mu}\sqrt{\frac{15 \mu}{\epsilon}}\approx 38$, which is a typical value for current two-phase simulations of turbulent flow~\citep{Loisy2017, Elghobashi2019}. Convergence of the statistical properties of the flow with respect to numerical resolution is achieved and is shown with simulations performed at lower maximum refinement. Note that while the flow is statistically stationary and equivalent for the various resolutions, because of the chaotic nature of the flow and the forcing, each realization presents differences one from each other. 

Figure \ref{figturb}b shows the statistical properties of the turbulent flow once the stationary state is reached. We characterize the fluctuations using the second order structure functions in the longitudinal $D_{LL}(d)$ and in the transverse direction $D_{NN}(d)$, defined as :
\bb
D_{LL}(d) &=& \frac{1}{3} \sum_i \langle \left(u_i(\bs r,t) - u_i(\bs r+ d {\bf \hat r_i},t) \right)^2 \rangle \\
D_{NN}(d) &=& \frac{1}{6} \sum_{i \neq j} \langle \left(u_i(\bs r,t) - u_i(\bs r+ d {\bf \hat r_j},t) \right)^2 \rangle,
\ee
for homogeneous and isotropic flows, with ${\bf \hat r_i}$ the unit vector along the $i$ direction. The transverse structure function is compensated by its scaling for a homogeneous and isotropic flow $(d\epsilon)^{2/3}$, and we indeed observe a plateau value close to $C=2$~\citep{Pope_book}. The relation $D_{LL} = 3/4 D_{NN}$ is also verified by representing the compensated longitudinal structure function $4/3 D_{LL}(d) (d \epsilon)^{-2/3}$. The inertial range is obviously quite limited due to the relatively coarse resolution, but the turbulent flow at the scale of the bubble to be injected is reasonable, and the bubble radius lies within the inertial range. The quantity of interest for the bubble deformation is the spherical velocity increments, which are defined on a sphere of radius $R$ by :
\bb
\delta u_S(R,\theta,\varphi) &=& {\tilde {\bu}}({\bf R},t) \cdot \er \\
\tilde {\bu}({\bf R}) &=& \bu - \iiOn \bu({\bf R},\theta,\varphi,t).
\label{sphere_inc}
\ee
From homogeneity and isotropy, the statistical properties of $\delta u_S(R)$ shall only depend on $r$. Figure~\ref{figturb}c shows the velocity increment $\delta u_S$ over a sphere of radius $R/\eta_k=32$, displaying an example of the broad range of forcing scales the bubble sees and feels in the flow. The ensemble average value $\delta u'_S(R) =\sqrt{ \langle (\delta u_S(R))^2 \rangle}$ compensated by $(\epsilon d)^{2/3}$ is represented in red in figure~\ref{figturb}b. We observe $\delta u'_S(R) = D_{NN}(d/2)$ for $Re_\lambda = 38$ in the entire inertial range. The statistical properties of the spherical increment and their link with surface deformations will be further discussed in section~\ref{DNS}. \\

Once the statistically stationary regime is reached, the temporal recording of the velocity field is stored from $20$ to $60 \tau$. Different instants are used as initial times for numerical simulations of bubble deformation and break-up. A central sphere of radius $R_0$, diameter $d_0$ and density $\rho_g = 1/850 \rho_\ell$ is placed in the periodic box, and the flow in the inner phase is initially set to zero. The bubble diameter is located in the inertial range. For $Re_\lambda = 38$ we have $d_0/\eta_K = 17.6$, $d_0/\lambda = 1.49$ and $d_0/L = 0.13$ where $\eta_K = (\nu^3/\epsilon)^{1/4}$ is the Kolmogorov length scale, $\lambda$ is the Taylor microscale $\lambda = \sqrt{15 \nu u'^2/\epsilon}$ for homogeneous and isotropic turbulence and $L$ is the box size. The level of refinement around the interface is $N=9$ for the majority of the numerical runs, and convergence tests at level $N=10$ have been performed. The results on bubble deformation by the turbulent flow presented here are independent of the resolution between refinement levels $N=9$ and $N=10$ as shown in the Appendix. These resolutions correspond to 70 and 140 points across the initial bubble diameter, which is comparable to the resolution successfully used to resolve rising motion and path instability of bubbles using the same numerical methods \citep{cano2016paths}.

In a turbulent flow, the critical Weber number, or break-up threshold, is defined in a statistical sense; therefore the probability of breaking does not vanish for Weber number immediately smaller than $\We = \We_c$. To determine the critical Weber number, we perform ensemble of simulations by using different initial times from the turbulence precursor simulation (each precursor typically spaced by $1-3t_c$). We do not observe any breakup at $\We = 1.5$ while running the simulations up to $20t_c$ for more than 10 runs with different initial conditions. For $\We=3$, we observe break-ups a little over 50\% of the time by $20t_c$ for an ensemble of 35 runs with different initial conditions. It is clear that given the broad lifetime statistics of a bubble in these conditions close to stability, the percentage of cases that do not break might change if the simulations are run for longer times. For $\We=6$, and above, all bubbles break within a few $t_c$. As such we consider $\We_c=3$ as the critical Weber number in our configuration and this value is within the variation of experimental measurements discussed in the literature.

\begin{figure}
\centering
\includegraphics[width=\linewidth]{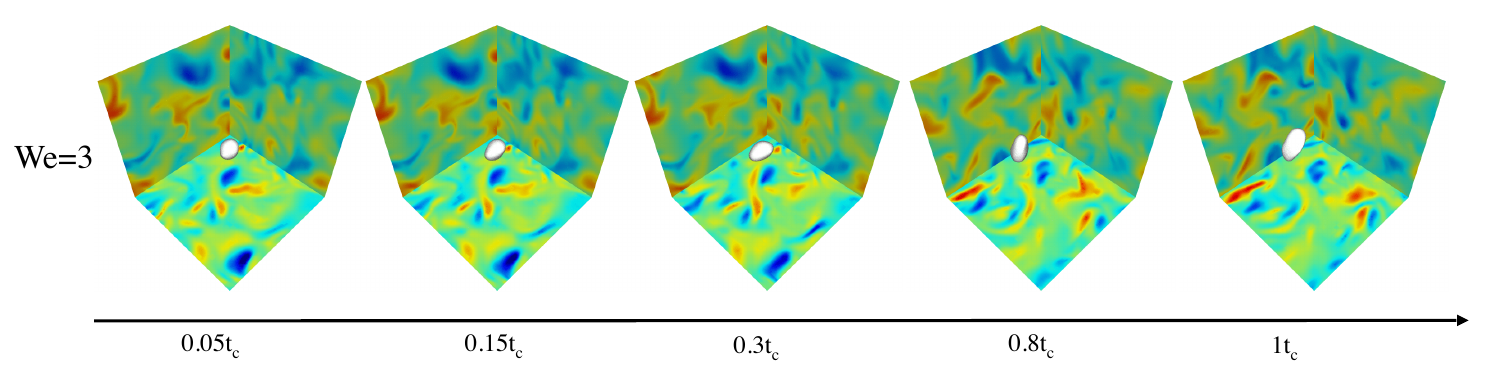}
\caption{Snapshots of a bubble for one run at $\We = 3$, with the interface in white and each background plane showing one component of the velocity. The bubble is injected in the center of the turbulent domain at $t=0$ with no velocity and starts to deform quickly. \review{During the first eddy turnover time, no break-up event occurs, but we observe strong and erratic deformations over time as well as advection by the turbulent flow.}}
\label{figbubble}
\end{figure}

The Weber number, eq. \ref{We}, is defined considering the temporal average of the turbulence dissipation rate, and we vary the Weber number by changing the surface tension. Starting from a sphere, we study the growth of the surface deformation until either the saturation of bubble deformation for stable bubbles or the breaking for unstable cases. We perform simulations for a wide range of Weber number, $1.5 < We < 45$. The critical Weber number is found \review{at} $\We_c=3$, \review{for} which about half of the runs do not exhibit bubble break-up after 20$\tau$. We will discuss simulations where for small Weber numbers ($\We \leq 1.5$) no break-up occurs, while for large Weber number ($\We \geq 3$) we will analyze the dynamics before the first break-up. We perform an ensemble of simulations of 10 to 15 runs for each We number, using various stored initial conditions from the precursor flow field, leading to a total of about 80 simulations.

Figure \ref{figbubble} shows an example of a bubble evolving in the turbulent flow field in the vicinity of the instability threshold at \We=3, during the first eddy turn over time at the bubble scale $t_c$. The bubble is initially spherical and quickly deforms for $t<0.5 t_c$, and then starts to exhibit erratic oscillations. These deformations do not lead to break-up, and occur together with advection by the turbulent background flow. 

\subsection{Computing the bubble deformation through a Voronoi decomposition}

\begin{figure}
\centering
\includegraphics[width=\columnwidth]{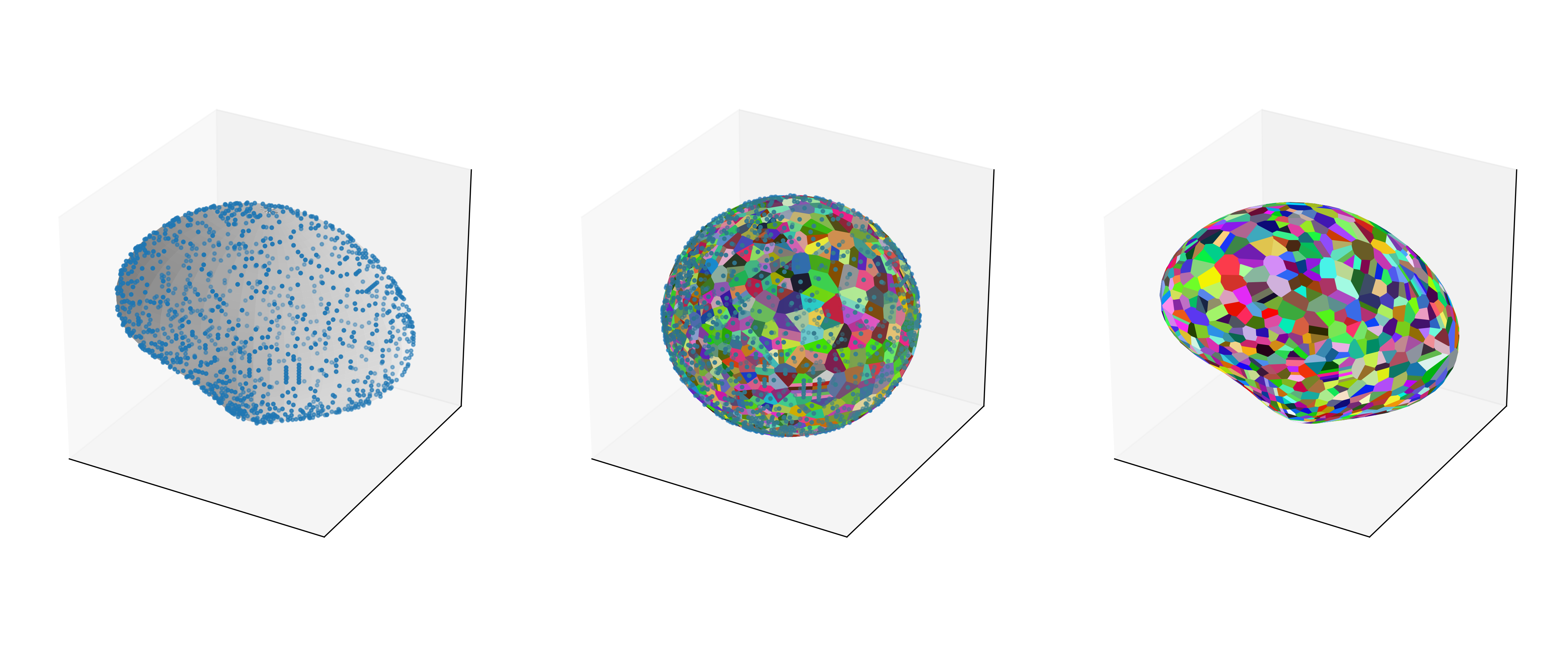}
\caption{Illustration of the interface tesselation using spherical Voronoi diagram. a) Bubble deformed by the turbulent background (one case at $\We = 3, t/t_c = 0.22$). The output data points are unevenly distributed at the interface. The bubble surface in grey levels has been reconstructed using a two dimensional spline interpolation. b) Projection of each point on the unit sphere, and computation of a spherical Voronoi diagram. Each polygon corresponds to the region associated to a single data point. c) Projection of the Voronoi diagram onto the initial shape bubble using the spline interpolation. The weight of each datapoint is given by the associated polygon area.}
\label{figvoronoi}
\end{figure}

The interface sampling points $\bf r_i$ are by essence non-uniformly spread, as more points are dynamically added in regions of larger curvature due to the adaptive algorithm. To compute averages accurately over the bubble interface, we then construct a surface scheme using a spherical Voronoi decomposition~\citep{Voronoi_1908}. The bubble surface is partitioned into regions close to each of the points. \review{For each region located around the point $\br_i$, we associate a region area $A_i$ and a corresponding solid angle $\Omega_i = A_i/r_i^2$. We evaluate the surface integral of any function $f$ by :
\bb
\iiOn f(\br) = \frac{1}{4 \pi}\sum_i f({\br}_i) \Omega_i,
\ee
\review{where $\sum_i \Omega_i = 4 \pi$}. The computation of the region areas $A_i$ and the corresponding solid angles $\Omega_i$ is illustrated in figure~\ref{figvoronoi}. The $\Omega_i$ are computed in three steps. We first project all interface points on the unit sphere (step 1). We then compute the region locations using a spherical Voronoi algorithm based on a robust Delaunay triangulation~\citep{Caroli_2009} (step 2). The area $A_i$ of each convex polygon is eventually estimated using the shoelace formula (step 3). On the unit sphere, the $A_i$ and the $\Omega_i$ are identical}. An example of our Voronoi decomposition is shown in Figure~\ref{figvoronoi}b on the unit sphere, and in Figure~\ref{figvoronoi}c after projection on the initial bubble shape. The numerical error on the total solid angle $\sum_i \Omega_i$ is less than 0.1$\%$.

To analyse the interface dynamics using Eq.~\ref{alm_eq1}, we introduce the spherical harmonics $\Ylmf$. The numerical estimate of the harmonic coefficient $\alm$ is obtained by a sum over the interface points $\br_i$ weighted by their local solid angle $\Omega_i$ :
\bb
\alm = \sum_i \Omega_i \zeta(\br_i) \Ylm (\theta_i,\varphi_i)
\ee
where $(r_i, \theta_i,\varphi_i)$ are the spherical coordinates of $\br_i$. We have first checked the orthonormal property of the spherical harmonics $\iiOn \Ylm \cdot Y_{(\ell',m')} = \delta_{\ell,\ell'} \delta_{m,m'}$ for $\ell \in [0,10]$ and all corresponding $m$ values on a set of 900 points, typical of a surface bubble sampling, and randomly spread on a unit sphere. The typical error is 0.1\% for $\ell < 5$, and increases for higher $\ell$ values. The harmonic decomposition has been tested on synthetic shapes, \textit{i.e.} a set of 900 points randomly spread on an interface of known spherical decomposition. The \review{relative error} on the harmonic coefficient is about $0.5\%$ of the largest non-zero harmonic coefficient for $\ell \in [1,5]$. In practice, the amplitude decreasing rapidly with $\ell$, an accurate estimate of the harmonics coefficient is limited to $\ell \leq 5$ for the interface deformation and around $\ell \sim 10$ for the velocity field.

The bubble center position evolves with time, the interface being advected by the turbulent background flow. To analyse specifically the surface deformations, we look for the centre position $\br_c$ of the bubble frame of reference, for which $a_{1,m} = 0$ for $m \in \{-1,0,1\}$, as defined in section 2. The centre position $\br_c$ is computed recursively as follows. At each step, we compute the Voronoi diagram and the associated spherical harmonic functions $Y_{1,m}$ using the regions area $A_i$. Then, since each function $Y_{1,m}$ presents a symmetry of revolution, a one dimensional gradient descent on each function is sufficient to find the position center that minimizes each coefficient $a_{1,m}$. For bubble interfaces with single valued radial distance $r_i(\theta,\varphi)$, the gradient descent indeed converges to a centre position $\br_c$ for which the three mode 1 coefficients $a_{1,m}$ vanish.

\subsection{Deformation dynamics for stable conditions: bubble deformation and temporal evolution of spherical modes}

\begin{figure}
\centering
\includegraphics[width=\columnwidth]{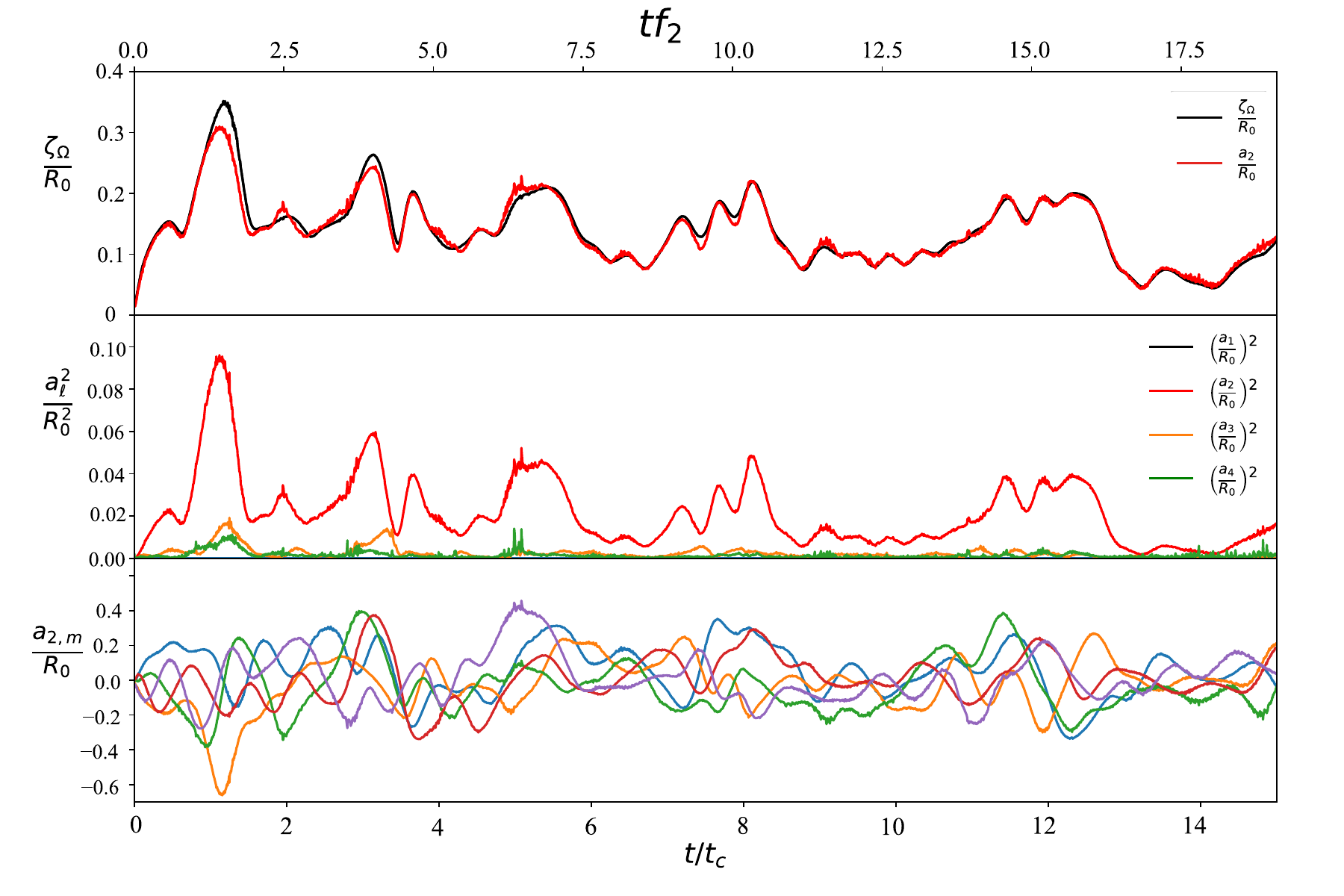}
\caption{(a) Total deformation $\zO/R_0$ as a function of time, together with the rms amplitude of the mode 2 oscillations. (b) Energy $a_\ell^2= \sum_{m=-l}^{m=l} a_{l,m}^2$ of the spherical harmonics for modes $\ell=1$ to $\ell=4$ for one run at $\We = 3$. Mode 1 energy vanishes by definition. Mode 2 appears as the most energetic mode, in agreement with experimental literature, with oscillation close to $f_2 = \omega_2/(2 \pi)$. As $\ell$ increases, the energy contained decreases sharply. (c) Temporal evolution of the amplitude coefficients $a_{2,m}$ for mode $\ell=2$, and $m=-2$ to $m=+2$. All modes $m$ oscillate with similar amplitude and frequency.}
\label{figharm}
\end{figure}
We first describe the temporal evolution of the coefficients $\alm$ for a single run at $\We$=3, as an illustration of the analysis performed on each simulation. Several modes $\alm$ of the same $\ell$ and different $m$ values are associated to the same oscillation frequency (see Eq.~\ref{alm_eq1}), therefore we introduce the global coefficient $a_\ell$ describing the energy contained in each mode $\ell$, defined by :
\bb
a_\ell^2 = \sum_{m=-\ell}^{\ell} \alm^2,
\ee
where $a_\ell$ is then positive by convention. The global surface deformation is obtained by a sum over all modes $\ell$. We recall the expression of the root mean squared deformation $\zO$: $\zO^2(t) = \sumlm \alm^2(t)$. Figure \ref{figharm}a shows the time evolution of $\zO/R_0$ together with the total amplitude of the mode 2, $a_2/R_0$. We observe a rapid linear rise, consistent with the prediction for $t\ll t_2,t_c$, before oscillations and saturation of the total deformation around $\zO/R_0 \approx 0.15$. As shown in the figure, most of the deformation comes from the mode $\ell=2$. This first observation is in agreement with various experimental studies on large bubbles immersed in a turbulent flow, which have reported the dominance of mode 2 deformation in bubble dynamics~\citep{Risso_1998,Ravelet_2011}. In practice, a sum over the first three modes $\ell= 2,3,4$ estimates the amplitude deformation $\zO$ within less than 2\% of error as long as $\zO/R_0 < 1$ for all cases. It confirms the predominance of the first modes of oscillation, and validates the spherical decomposition approach. In the following, we focus on the modes $\ell = 2,3$ and 4 which contribute mostly to the surface deformation.

\begin{figure}
\centering
\includegraphics[width=\columnwidth]{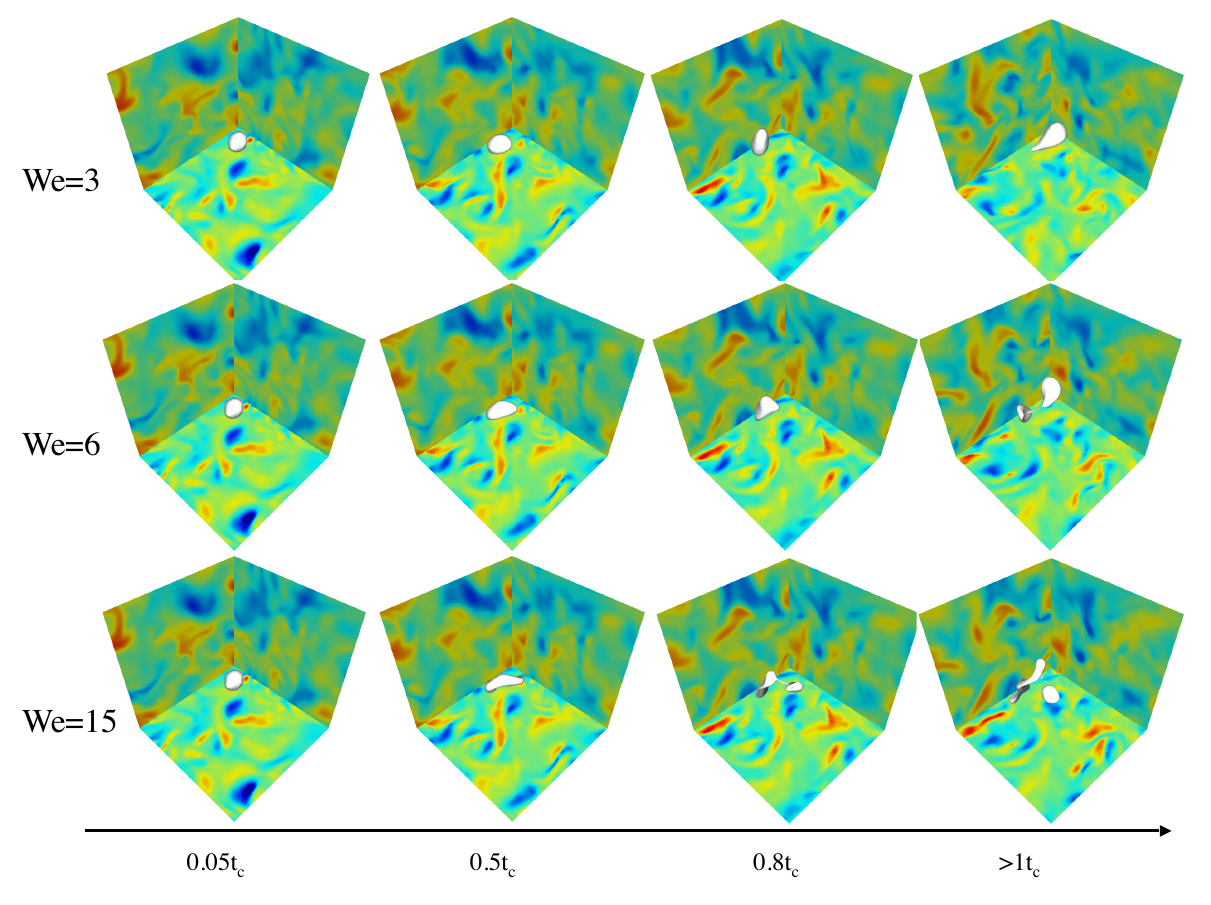}
\caption{Bubble deformation and break-up as time progresses for the same initial conditions but increasing Weber number. Time increases from left to right while We increases from top to bottom. At $\We$=3, this particular condition does not lead to break-up. At $\We$=6, break-up occurs for $t/t_c\approx 1.5$, while break-up occurs earlier for $\We$=15, at $t/t_c\approx 0.8$. The initial stage of deformation, for $t<0.5t_c$ appears very similar at all Weber numbers.}
\label{figmwe}
\end{figure}

Figure \ref{figharm}b shows the temporal evolution of $a_\ell^2$ for the first few spherical harmonics $\ell \in [1,4]$ for a single run. The modes $\ell=1$ have zero amplitude from the choice of the center position, and the modes $\ell=2$ dominate the surface energy deformation. The energy contained in modes $\ell=2$ oscillates with time, at a frequency close to $2 \omega_2$ and the energies in the higher order modes $\ell=3, 4$ and $5$ remain smaller at all times. Figure \ref{figharm}c shows the temporal evolution of each individual harmonic coefficient $a_{2,m}$ for different $m$ components. All 5 coefficients appear to oscillate with comparable amplitude and with frequencies close to $\omega_2$. 

\subsection{Deformation dynamics for increasing Weber number and unstable conditions}

For one particular configuration of the turbulent flow, we investigate the role of the Weber number on the deformation by changing only the value of the surface tension. Figure \ref{figmwe} shows snapshots of the bubbles for increasing Weber number $\We$ = 3, 6 and 15, for the same initial turbulent conditions. For all cases, we observe early growth of surface deformation, while the turbulent background flow stays identical for all runs and independent of surface tension, validating the no-feedback hypothesis introduced in section 2. For all We, the short time deformation under the same turbulence conditions appears identical, in agreement with the theoretical description of the linear regime for $t\ll t_c$. For stable conditions (here We=3), the deformation rapidly saturates as previously described. For unstable conditions (here We=6 and 15), the deformations eventually lead to break-up at various times. The highest Weber number ($\We$=15) displays a break-up event relatively early at $t/t_c\approx 0.8$, while for conditions closer to the stability threshold, break-up occurs at $t/t_c>1$.

\begin{figure}
\centering
\includegraphics[width=\columnwidth]{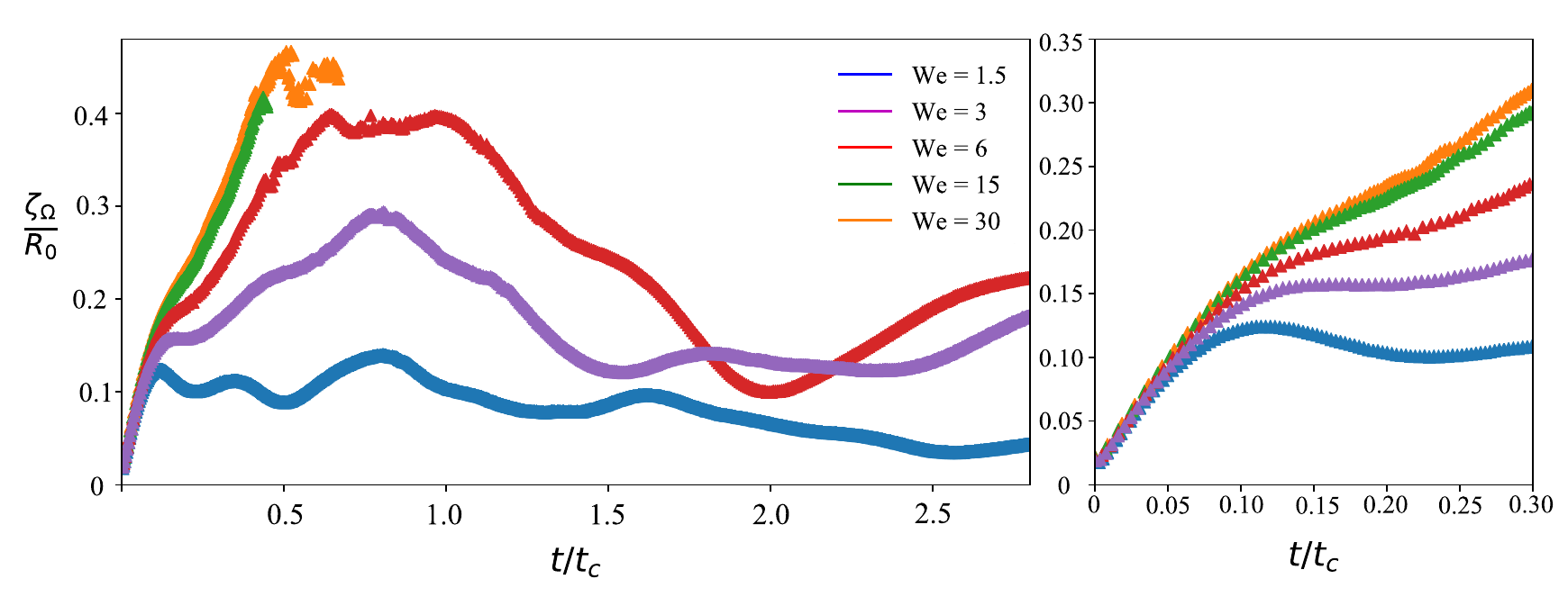}
\caption{(a) Root mean square surface deformation $\zO$ as a function of time $t/t_c$ for the same initial turbulent flow condition and increasing Weber number $\We$ = 1.5, 3, 6,  15, 30.  (b) Zoom-in on the early-stage dynamics. For $t/t_c<0.1$, a universal linear increase of the deformation is observed at all We number. Earlier saturation time is reached for higher surface tension forces (lower $\We$), leading to lower values of the saturated deformation amplitude.}
\label{figrms}
\end{figure}

A quantitative description of the Weber number influence for one particular run is given in Figure~\ref{figrms}. We compute the total deformation $\zO$ starting from the same initial condition of the flow configuration, and increasing values of the Weber number $\We$ = 1.5, 3, 6, 15, 30 and 45. Figure \ref{figrms} shows the deformation $\zO/R_0$ as a function of time $t/t_c$. We observe a universal rapid linear growth, followed by a saturation that depends on the Weber number. The rapid linear growth is independent of the Weber number for $t/t_c<0.1$, as expected from Eq.~\ref{alm_lin}. As time passes, the curves diverge from each other, and the lower Weber number curves exhibits earlier sub-linear growth. The saturation leads to long time oscillations at the low Weber numbers, namely $\We$=1.5, 3 and 6 during several eddy turnover times, with increasing amplitude as the Weber number increases. For larger Weber number (here $\We$=15 and 45), the bubble experiences break-up during the first eddy turn over time, when surface deformation reaches $\zO/R_0 \approx 0.74$. \reviewsecond{Note that the deformation at break-up is estimated from a fit of the bubble interface by an ellipsoidal shape at the break-up time, as the Voronoi decomposition fails at large deformation, significantly before break-up.}

\subsection{Ensemble-averaged deformations at one Weber number: linear growth and saturated regime.}

\begin{figure}
\centering
\includegraphics[width=1\columnwidth]{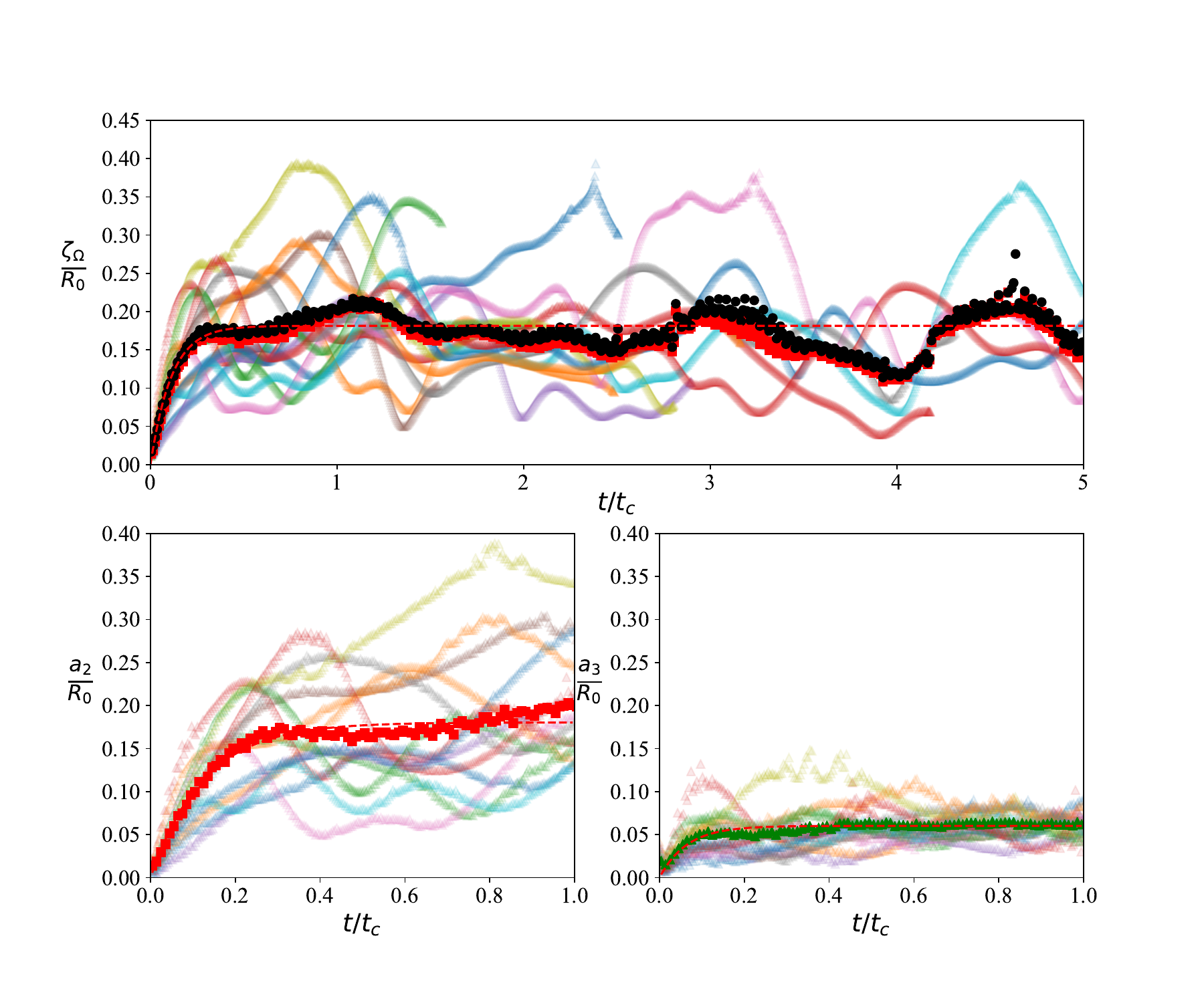}
\caption{a) Individual realisations of $\zO$ at $\We$ = 3, as a function of the dimensionless time $t/t_c$ in transparent colors. Ensemble average $\langle \zO \rangle$ of the surface deformation is superimposed (black full symbolds), together with the ensemble average mode 2 amplitude (red full symbol). b) Coefficient $a_2$ as a function of dimensionless time $t/t_c$ for $\We=3$ (individual realisations in transparent colors). Ensemble average $\langle a_2\rangle$ is superimposed (full red symbol). Fit by $A (1-e^{-t/t_{sat}})$ of the ensemble average $a_2$ in shown by the dashed lines. c) Coefficient $a_3$ as a function of dimensionless time $t/t_c$ for $\We=3$ and the equivalent exponential fit (individual realisations are transparent colors and the green full triangle is ensemble average $\langle a_3\rangle$).}
\label{figrmsE}
\end{figure}

To obtain ensemble average quantities, we analysed between 10 to 15 runs for each Weber number, starting from different turbulent flow configurations, which are obtained from different times of the precursor simulations. For each run, we perform the analysis described above, from the Voronoi decomposition to the computation of the surface deformation $\zO$ and the first harmonics coefficients $a_2, a_3$ and $a_4$. 
	
Figure~\ref{figrmsE}a) shows the temporal evolution of each individual realisation of $\zO/R_0$ performed at $\We = 3$, as a function of the dimensionless time $t/t_c$. The ensemble average $\avg{\zO}/R_0$ is superimposed in black, and follows a linear increase, as predicted by eq. \ref{alm_lin_avg}. The mode 2 amplitude $a_2/R_0$ is superimposed in figure~\ref{figrmsE}a) ($\textcolor{red}{\square}$), confirming in a statistical sense the predominance of the mode 2 of oscillations. The modes $a_2/R_0$ and $a_3/R_0$ follow the same trend, as shown in figure~\ref{figrmsE}b) and c), each coloured curve being an individual realisation. Again, at short time the linear regime predicted by eq. \ref{alm_lin_avg} is observed.

We observe a transition to a sub-linear regime for $\avg{\zO}$, $\avg{a_2}$ and $\avg{a_3}$ around $0.1 < t < 0.3$. This transition is well described by the theoretical expression derived in section~\ref{theory} (eq. \ref{complex_fit}) \textit{i.e.} $\propto t \sqrt{(1 - (\kappa t)^2)}$ with $\kappa= \omega_2/\sqrt{3}$ for $\avg{\zO}$ or $\avg{a_2}$, and $\kappa=\omega_3/\sqrt{3}$ for $\avg{a_3}$. The prefactor is related to the turbulence velocity field statistics $\ulm$, which will be evaluated in an upcoming section. In order to evaluate the saturation with time, we describe the temporal evolution of $\avg{\zO}$, $\avg{a_2}$ and $\avg{a_3}$ by the empirical expression $a_\ell^\infty \left (1 - e^{-t/t_{sat}} \right )$ with two fit parameters $a_\ell^\infty$ and $t_{sat}$. The early time evolution is proportional to $a_\ell^\infty/t_{sat}$, while the characteristic time of saturation is given by $t_{sat}$. Eventually, we perform for each ensemble average mode $\avg{a_\ell}$ a linear fit $a_\ell = s_\ell t$ in the early time evolution ($t<t_{sat}/2$) by a simple linear model, which gives an accurate measure of the growth velocity $s_\ell$.

	
\subsection{Ensemble averaged deformations for increasing Weber number}
\begin{figure}
\centering
\includegraphics[width=\columnwidth]{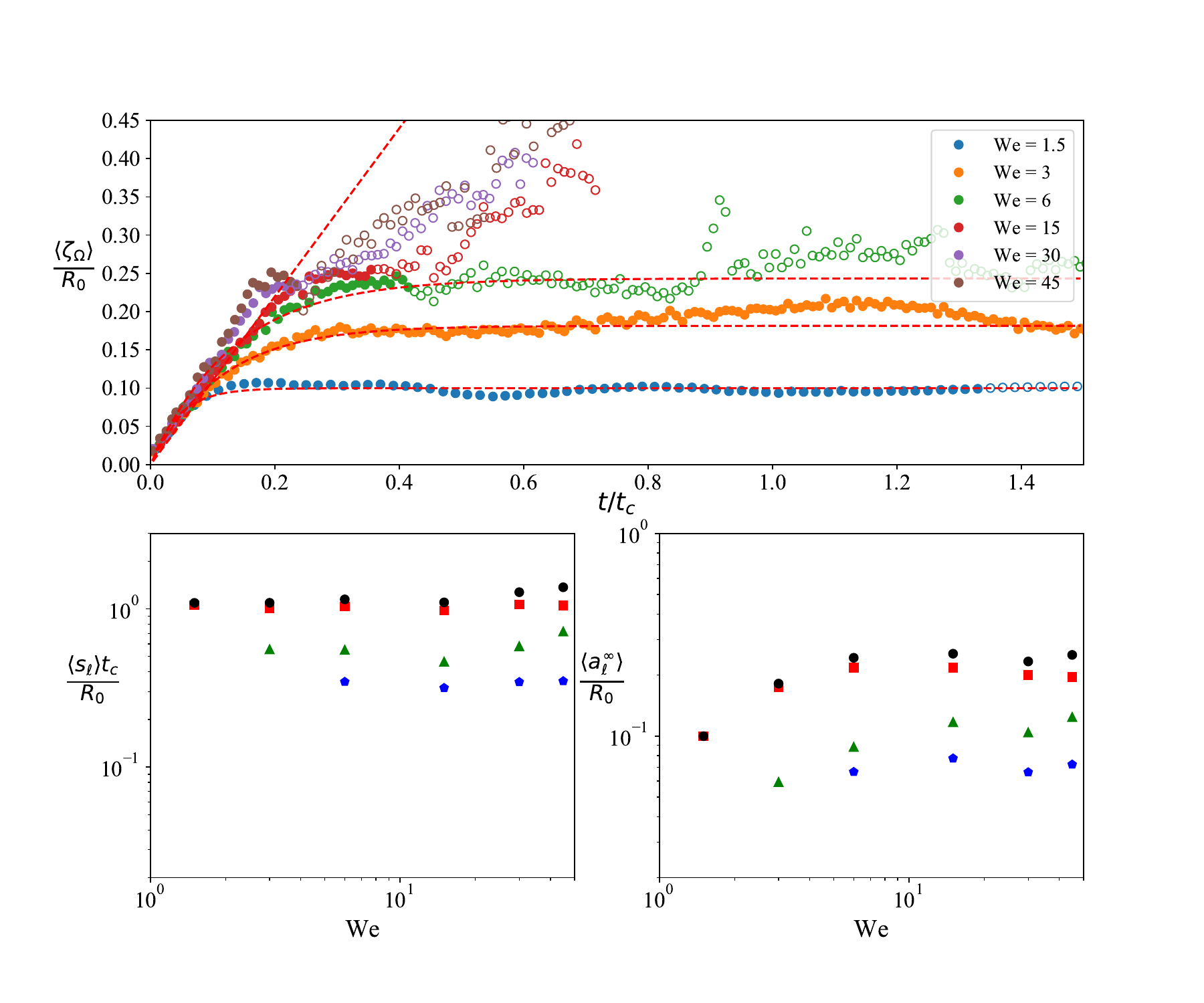}
\caption{a) Ensemble-averaged deformation $\langle\zO\rangle$ as a function of time for increasing We number. For the lower $\We$, a fit by $A (1-e^{-t/t_{sat}})$ captures the transition from linear growth to saturation, with saturation values increasing with We. Open symbols are used once half of the bubbles in the ensemble have broken. b) Slope at the origin as a function of the Weber number for $\langle\zO\rangle$, $\langle a_2 \rangle$, $\langle a_3 \rangle$, $\langle a_4 \rangle$. The slope at the origin for $\langle\zO\rangle$ is independent of We. c) Amplitude of saturation of the global deformation and modes as a function of We.}
\label{figensemble}
\end{figure}

Figure~\ref{figensemble} summarizes the surface deformation growth for increasing We number. Figure~\ref{figensemble}a) shows the evolution of $\langle\zeta_\Omega \rangle$ averaged over at least 10 runs by Weber number, as a function of the Weber number. The evolution is shown in solid symbols until half of the bubbles have broken, and in open symbols later. At early time, we show the existence of a universal regime, with a linear increase in time of the deformations independent of the Weber number We. Similar results are observed when considering the spherical harmonics amplitude $\avg{a_2}$, $\avg{a_3}$ and $\avg{a_4}$. The dimensionless growth velocity $s_\ell t_c/R_0$ at short time, obtained from the linear fits, is shown in figure~\ref{figensemble}b) as a function of the Weber number for $\avg{\zeta_\Omega}/R_0$ ($\circ$), $\avg{a_2}$ (red square), $\avg{a_3}$ (green triangle) and $\avg{a_4}$ (blue pentagon). For the low Weber numbers We=1.5 and We = 3, only the modes that are significantly above the noise level have been represented, excluding modes 3 and 4 at We = 1.5, and mode 4 at We = 3. The growth velocity $s_\ell$ depends on the mode $\ell$, but is independent of Weber number, as expected from Eq.~\ref{alm_eq1} and \ref{alm_lin_avg}. 

The departure from linear growth that occurs at a time increasing with Weber number is compatible with the prediction given the second mode reduced period $t_2$. Eventually, the saturation can be measured using the exponential fit for the lowest Weber number (We= 1.5, 3 and 6) where a saturation is clearly visible. For higher Weber number, we consider the value of the linear fitting model, evaluated at the time of the first break-up. The saturation time for $\We < 10$ and the earliest break-up value for $\We > 10$ is shown in figure~\ref{figensemble}c). We observe a continuous growth with Weber number, premising the likelihood of bubble break-up. At higher Weber number, we also remark that the relative amplitudes of modes 3 and 4 are increasing for both the growth velocity $s_\ell$ and the saturation $a_\ell^\infty$. At high We number, the higher order modes of deformation in the global dynamics become more important and are likely to influence the break-up geometry that the bubble will experience. \review{Note that treating the saturation at intermediate Weber number would require considering the role of non-linear and viscous effects. Note also that for $\zeta_\Omega/R_0> 0.25$, a significant part of the bubbles have broken, and the mean deformation observed when considering all break-up events is $\zeta_\Omega/R_0\approx 0.74$, which is compatible with the observations on areas of deformation at break-up made by \citet{Risso_1998}.}

Having access to the full velocity flow, we can now relate the surface deformation growth parametrized by $a_\ell$ and $s_\ell$ to the statistics of the surrounding turbulent flow.

\subsection{Statistics of velocity increments on a sphere predict the linear growth of deformation}
\begin{figure}
\centering
\includegraphics[width=\columnwidth]{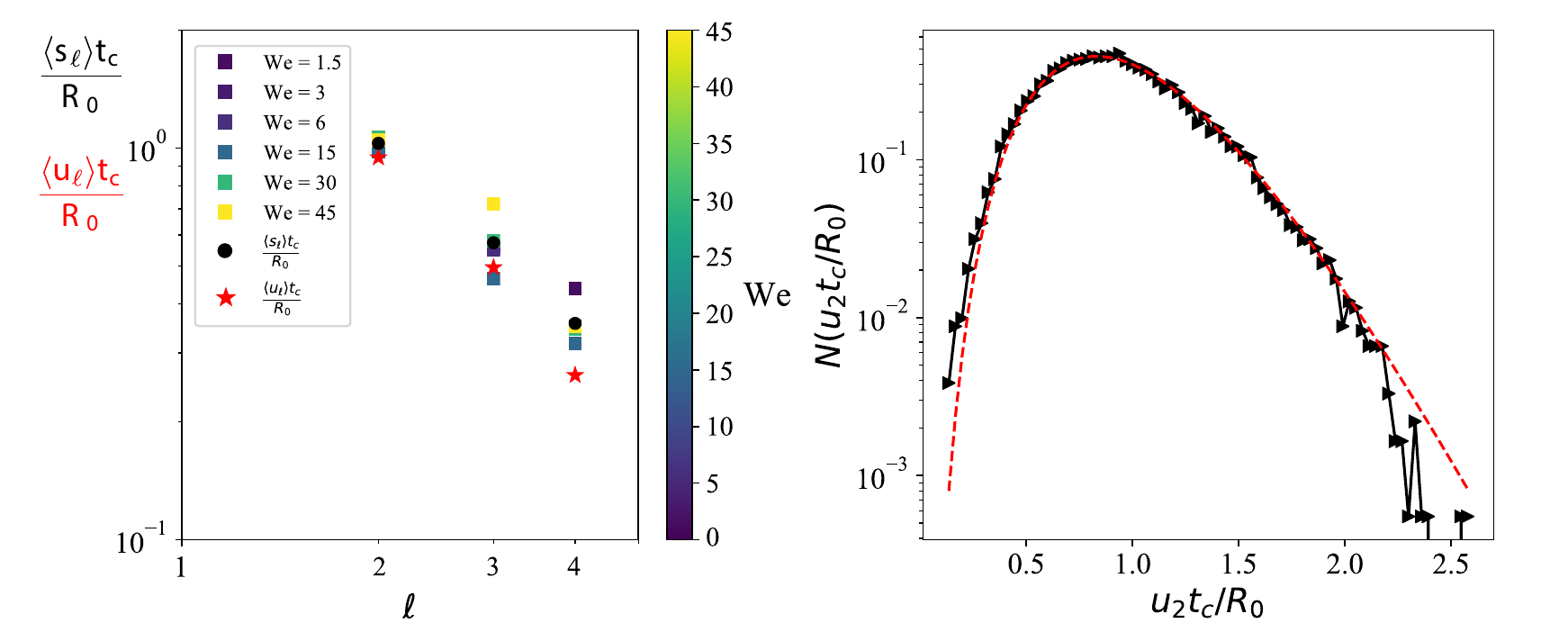}
\caption{(a) Ensemble average of the spherical harmonics growth rate $\langle s_\ell \rangle R_0/t_c$ for $\ell =2, 3$ and 4, together with \reviewsecond{the corresponding spherical harmonics flow coefficient $\langle u_\ell\rangle t_c/R_0$ of the turbulent velocity fluctuations at the bubble scale, shown as red pentagon. Black circles are $\langle s_\ell \rangle t_c/R_0$ averaged for all We, while the colored squares correspond to the We number scale (see colorbar).} The theoretical prediction $a_\ell = u_\ell t$ from section 2 is well verified, with velocity fluctuations providing a quantitative prediction of the bubble deformation. The  interface deformation statistics at early times can thus be extracted from velocity fluctuations at the bubble scale. (b) Probability distribution of the velocity fluctuations $\langle u_2 \rangle$ corresponding to the statistics of velocity fluctuations on the bubble sphere responsible for mode 2 deformation. A fit by a $\Gamma$-distribution of expression $f(x,k,p)=x^{k-1} e^{-x/p}/\Gamma(k) p^k$ is superimposed (red dashed line) with $k=7.5$ and $p=0.135$.}
\label{figalul}
\end{figure}

From the theory developed in section 2, the pre-factor of the linear growth regime of the spherical harmonics mode amplitude shall be related to the statistics of the turbulent flow. Indeed, the growth of the spherical harmonics coefficients $a_{\ell,m}$ is linked to their counterpart in the turbulent fluctuations, $\lm{u}$ and $\lm{\pi}$ in equations ~\ref{alm_eq1}, \ref{alm_lin_avg} and \ref{complex_fit}. It bears significant importance from a practical point of view, as measuring velocity statistics in an experimental or natural turbulent flow is relatively accessible while measuring deformation properties on a bubble interface is much more challenging. We recall the description of the spherical increments, $\delta u_S$, which can be decomposed in the spherical harmonics base, $\delta u_S = \sumlm \lm{u} \Ylmf$, where $\lm{u}$ depends on the radius $r$ of the sphere. 

Using the velocity flow in the DNS, we computed the ensemble average $\avg{u_\ell}$ of the harmonic coefficients of $\delta u_S(r)$ for $r=R_0$, using the same procedure as for the $a_\ell$. The dimensionless coefficients $\avg{u_\ell}t_c/R_0$ are shown in figure~\ref{figalul}a) as red pentagons, for $\ell = 2, 3$ and $4$ and correspond to the intensity of velocity fluctuations at the bubble scale for specific modes $\ell$. The coefficients $s_\ell t_c/R_0$ obtained from the processing of surface deformations are superimposed in black circles and correspond to the intensity of deformations of the bubbles for the corresponding modes $\ell$. For modes $\ell = 2$, 3 and 4, we find a quantitative agreement between $\avg{\alm}$ and $\avg{\ulm} t$, as expected from section 2, eq. \ref{alm_lin_avg}. For $\ell = 4$, we observe a slight difference for $\ell = 4$ with $s_4 > u_4$, which can be attributed to the limit of resolution of the harmonics coefficient computation on the bubble deformation for higher $\ell$ values.

Beyond the equality of the ensemble average values, the prediction made in section 2 shall be valid for each individual realisation. This equality can be used to infer the full statistics of $\alm$ at short time, using the statistics of the $\ulm$ in the background flow.


\begin{table}
\centering
\begin{tabular}{p{1.5cm} p{1cm} p{1cm} p{1cm} p{1cm} p{1cm} p{1cm} p{1.5cm} p{1cm}}
We= & 1.5 & 3 & 6 & 15 & 30 & 45  &\vline ~Re$_\lambda$ = & 38 \\
\hline
$\avg{s_2} R_0/t_c$ &1.06 & 1 & 1.03 & 0.93 & 0.98 & 0.95 & \vline ~$\avg{u_2}t_c/R_0$ & 0.95\\
$\avg{s_3} R_0/t_c$ & 0.56 & 0.55& 0.54& 0.44 & 0.51 & 0.64 & \vline ~$\avg{u_3}t_c/R_0$ & 0.50\\
$\avg{s_4} R_0/t_c$ & 0.44&0.35&0.34&0.31& 0.33 & 0.35 &\vline ~$\avg{u_4}t_c/R_0$ & 0.26\\
$s_2' R_0/t_c$& 0.41& 0.36 &0.33&0.32&0.33&0.27 &\vline ~$u_2' t_c/R_0$ & 0.34\\
$s_3' R_0/t_c$&0.23& 0.24 &0.23&0.15&0.23&0.27&\vline ~$u_3' t_c/R_0$ & 0.17\\
$s_4' R_0/t_c$&0.08 & 0.06 &0.08&0.08&0.10&0.09&\vline ~$u_4' t_c/R_0$ & 0.09\\
\hline
\end{tabular}
\caption{Values of the coefficients shown in figure 10 and 11. Ensemble average of spherical harmonics growth rate $\langle s_\ell \rangle R_0/t_c$ for $\ell =2, 3$ and 4. The corresponding spherical harmonics amplitude $\langle u_\ell\rangle t_c/R_0$ of the turbulent velocity fluctuations at the bubble scale is also provided, computed for Re$_\lambda$=38. \label{Table2}}
\end{table}

Figure ~\ref{figalul}b) shows the probability density function of the dimensionless mode 2 $u_2 t_c/R_0$ at the bubble scale $R_0$. In contrary to the two-points velocity increment $D_{LL}(R_0)$, the pdf of $u_2$ in the inertial range does not exhibit large tails, and 95\% of $u_2$ values lie in the range $[0.5 \avg{u_2},2 \avg{u_2}]$. The difference between the large tails of the pdf of $D_{LL}$ at $d/\lambda = 1.5$ and the short tails of $u_2$ can be attributed to the spatial average operation on the sphere, which smooths out all the intermittent structures at a scale smaller than the bubble, and suggests that the flow intermittency has a limited influence on the bubble deformation in the inertial range. Considering a positive definite quantity, a fit of the probability density function by a $\Gamma$-distribution of expression $f(x,k,p)=x^{k-1} e^{-x/p}/\Gamma(k) p^k$ gives $k=7.5$ and $p=0.135$. 


\section{Implications for bubble lifetime statistics}


\begin{figure}
\centering
\includegraphics[width=\columnwidth]{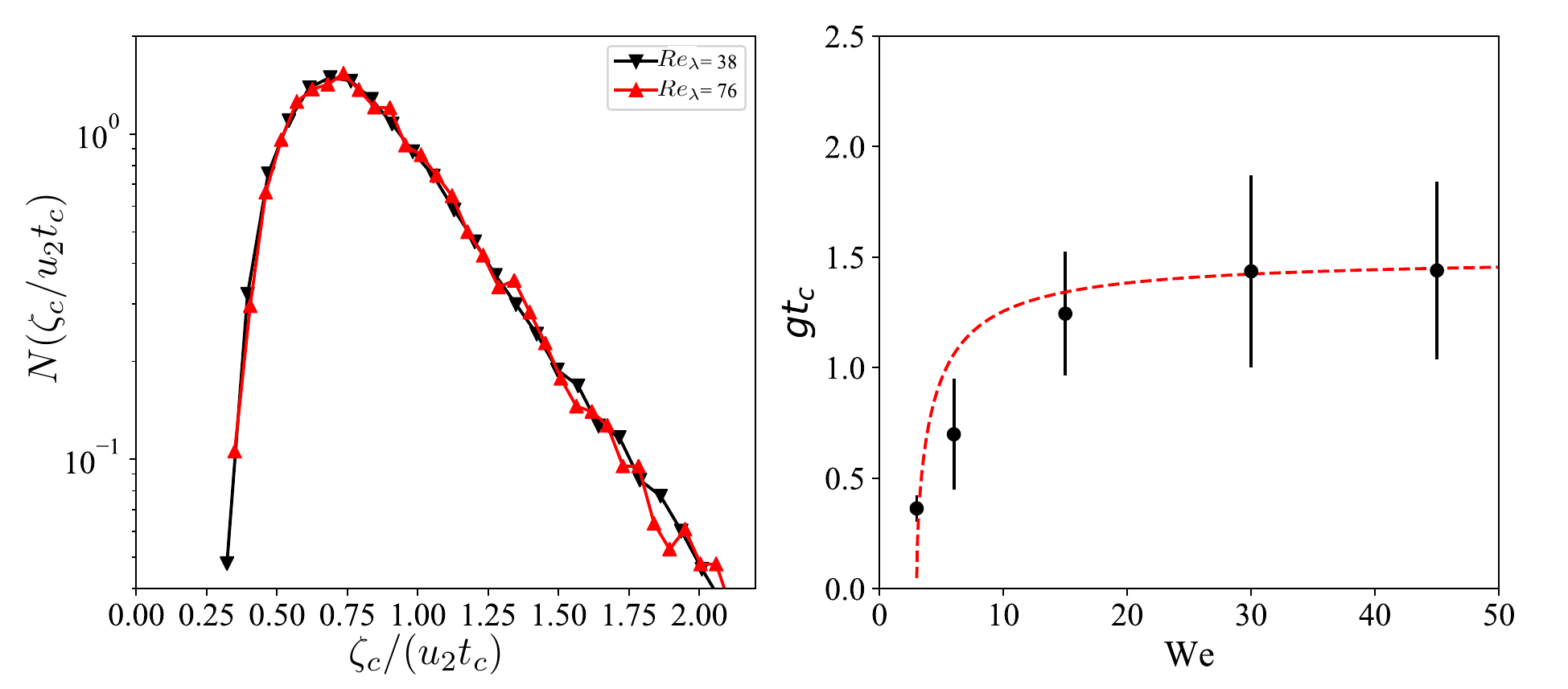}
\caption{\review{(a) Estimated probability distribution function of bubble lifetime in the limit of large Weber number, obtained from $N(T/t_c) =  N(\zeta_c/u_2 t_c)$ (eq. 4.1), and valid at high Weber number, \textit{i.e.} $\We \approx > 30$, using $\zeta_c = 0.74R_0$. Mean and rms values are given in table 3. (b) Inverse of the bubble lifetime $t_c/T$, and its standard deviation, similar to the bubble break-up frequency, as a function of the We number. The mean values for 10 to 20 simulations for each Weber number are used, and the error-bar corresponds to the standard deviation in the lifetime. Dashed line corresponds to the model proposed by \citet{Lasheras1999,Martinez2010}, $t_c/T\propto \sqrt{1-\We_c/\We}$. At high We number, the bubble lifetime can be predicted from the turbulence statistics, and given by $\zeta_c/u_2 t_c$}}
\label{fig_lifetime}
\end{figure}

\review{Given the linear growth of deformation with time and a growth rate given by the mode $u_2$, the pdf of $u_2$ can be used to evaluate the bubble lifetime distribution at high Weber number, when surface tension becomes negligible. In the limit of inertial break-up, with correlated velocity fluctuations The distribution of lifetime $N(T/t_c)$ at high Weber number is given by :
\bb
N(T/t_c) =  N \left ( \frac{\avg{\zeta_c}}{u_2 t_c} \right ),
\ee
where $\avg{\zeta_c}$ is the average critical deformation of $\zeta_\Omega$ at break up. $\zeta_c$ cannot be computed from the Voronoi decomposition of the interface, since the radius becomes multivalued at high deformation. To evaluate the critical deformation $\zeta_c$, we approximate the shape at break-up by an ellipsoid of the same volume than the initial bubble and whose longest axis corresponds to the maximum distance between two points on the bubble interface. From the large Weber number data ($\We \geq 15$), we find $\avg{\zeta_c} = 0.74$.
Figure~\ref{fig_lifetime}a shows the distribution $N(T/t_c)$ obtained from eq. 4.1, for a bubble of size $d_0/\lambda = 1.5$ using the statistics of the mode $u_2$ of the harmonic decomposition of $u_T$ for a Taylor Reynolds number $Re_\lambda = 38$ (the turbulence considered in the present paper). The lower lifetime bound is 0.1$t_c$, while the distribution shows exponential tails for large values of the bubble lifetime. We also compute the same predicted lifetime statistics for a turbulent flow obtained similarly but with higher Reynolds number, $Re_\lambda = 76$, and test values of $d_0/\lambda = 1.5$ (red line on fig~\ref{fig_lifetime}a) and $d_0/\lambda = 3$ and observe only little changes in the distribution. This suggests that the lifetime statistics are fairly insensitive to the Reynolds number for bubbles within the inertial range ($d_0/\lambda>1$).} \\

\review{The average bubble lifetime can be computed from the numerical runs for each Weber number from the break-up time of the initial bubble. Figure~\ref{fig_lifetime}b shows the inverse of the bubble lifetime $t_c/T$, similar to the break-up frequency introduced in the literature, for increasing Weber number. The errorbars correspond to the rms values $T_{rms}/t_c$. We observe that the bubble lifetime decreases with Weber number, and reaches a value independent of Weber number for $\We\gg \We_c$, here above We=30. We observe that our data can be reasonably well described by the model proposed by \citet{Lasheras1999,Martinez2010} (dashed line), $t_c/T = C_g \sqrt{1 - \We_c/\We}$., with the saturation value at high Weber number being the only adjusted parameter. We use $\We_c = 3$ and $C_g = 1.5$ is fitted to our numerical data, whereas \citet{Lasheras1999,Martinez2010} used $C_g = 0.673$ and $\We_c = 1$ to fit their data.}
\begin{table}
\centering
\begin{tabular}{p{1.3cm} p{1.3cm} p{1.3cm} p{1.3cm} p{1.3cm} p{1.3cm} p{3cm}}
 & We = 3 & We = 6 & We = 15 & We = 30 & We = 45 & \vline ~ Re$_{\lambda}=38$ \\
\hline
$\#$ of elements & 45 (13) & 39 (14) & 20 (8) & 20 (7)& 20 (7) & \vline ~  \\
$T/t_c$ &3.01& 2.15& 1.06& 0.90 &0.88 & $\vline ~\avg{\zeta_c}/(\avg{u_2} t_c)=0.91$\\
$T_{rms}/t_c$ &0.86&1.18&0.65&0.45&0.40& $\vline ~\avg{\zeta_c}/(u_2' t_c)=0.41$\\
$\avg{\zeta_c}/R_0$ & 0.25& 0.5& 0.68 & 0.85& 0.69 &
\end{tabular}
\caption{\review{Bubble life-time obtained as a function of Weber number, obtained from the DNS and from the turbulence statistics. The mean $T$ and rms $T_{rms}$ values of the bubble life-time obtained from ensembles of 20 to 45 simulations at increasing Weber number (number of elements indicated in first row). We also provide the computed mean deformation at break-up $\avg{\zeta}$. The number in parentheses corresponds to the number of elements used to compute the deformation at break-up $\zeta_c$ from the bubble deformation simulations. On the right of the table, we show the predicted mean and rms lifetime from the turbulence statistics, $\zeta_c/(\avg{u_2}t_c)$, for a constant deformation threshold $\zeta_c/R_0 = 0.74$, for Re$_{\lambda}=38$. Excellent agreement between the numerical values and the ones predicted by the turbulence statistics is observed at high Weber number. \label{Table3}}}
\end{table}

\review{Table 3 summarizes the bubble lifetime and the estimated lifetime from velocity statistics. The mean and standard deviation values of the inferred lifetime from the turbulence statistics can be compared with the bubble lifetime obtained from the direct numerical simulations for ensembles of 20 to 45 simulations. We observe excellent agreement between the inferred and simulated mean lifetime, as well as for the rms value of the distribution, considering a constant break-up threshold of $\zeta_c = 0.74 R_0$, inferred from the bubble deformation at break-up for high Weber number cases ($\We \geq 15$). The full distribution $N(T/t_c)=N \left ( \avg{\zeta_c}/(u_2 t_c) \right )$ could hence be a good estimate of bubble lifetime statistics in the limit of high Weber number.}
%

\section{Conclusion}

We have presented a theoretical framework for bubble deformation in a turbulent flow, by performing a spherical harmonics decomposition of the bubble deformation and deriving a general forced oscillator equation for these spherical harmonics modes, where each mode is forced by the corresponding turbulent fluctuations mode. We identify various regimes in the time evolution of the deformation, in particular a short time scale regime where deformations grow linearly with time, the pre-factor being given by the strength of the turbulent flow.

We perform direct numerical simulations of bubbles evolving in homogeneous and isotropic turbulent flow at increasing We number and verify the theoretical predictions. We observe the linear regime of deformation growth for individual events and ensemble averaged quantities and show that the pre-factor of the linear growth is indeed given by the statistics of the turbulence deformation at the bubble scale. We observe in the simulations that the level of deformation saturates, with a saturation time and level of saturation that depends on the Weber number, with lower Weber number saturating earlier and at lower deformation. At low to intermediate Weber number, the eigen-mode 2 dominates the deformation. At lower We number, a much broader lifetime distribution is observed, with break-up that can occur at much later time, as multiple eddies participate in the deformation and break-up, as described in \cite{Risso_1998}. The broad lifetime distribution can be interpreted as a consequence of a stochastic process with a threshold, which gives rise to long oscillation before a fluctuation of higher amplitude leads to break-up. Further stochastic analysis of the surface deformation fluctuations around the average saturation, accounting for non-linear and viscous effects, could provide a quantitative prediction of the lifetime distribution at intermediate Weber number.

\review{At high Weber number, break-up occurs within a single eddy turn-over time, and close to the linear regime of deformation. These findings have significant implications: the ability to predict bubble deformation and break-up time from the turbulence statistics is now possible, by extracting the turbulence statistics and computing the spherical harmonics decomposition of the corresponding mode 2 and 3. According to our theory and as shown in the final section, the statistics of such modes directly provide the bubble lifetime, if one assumes a maximum deformation threshold. Our data typically suggest $\zeta_\Omega/R_0 \approx 0.74$. The deformation also shows significant increase of higher frequency modes, which could be related at high Weber number to the appearance of much smaller child bubbles. A more elaborate model would require a better estimate of the deformation statistics at break-up.}

\reviewthird{Experimental work probing the full three dimensional turbulence, while challenging, could further test these results on bubble deformation, bubble lifetime and break-up in turbulence. Another prospect involves simulations at higher turbulent Reynolds number investigating the roles of the ratio between the turbulent length scales and the bubble deformation dynamics, as well as the local coupling between the turbulent flow in the dense phase and the bubble deformations. Similarly, this numerical framework could be used to explore the child bubble size distribution resulting from bubble break-up (see \citet{Riviere2021}) and its relationship with the turbulence statistics.}

\section*{Appendix : Convergence test}
We present a convergence test on the interface resolution, by running simulations for maximum interface refinement levels $N=9$ and $N=10$, using the same precursor simulation. As discussed in section 3, these resolutions correspond to 70 and 140 grid points per diameter, while the resolution of the turbulent flow remains the same, and correspond to the $Re_{\lambda}=38$ flow with a velocity refinement of level 7. The interface deformation $\zeta_\Omega$ as a function of time is shown in figure 12, for increasing Weber number. We observe quasi-identical results between the maximum levels 9 and 10, for all cases, for times of the order of the eddy turn-over time. This shows that the interface deformation is correctly resolved at the chosen resolution, and the results presented in the paper are independent of numerical resolution. It is important to remember that the increase in resolution here only applies to the interface resolution, while the resolution on the turbulent velocity field remains the same. However, we have independently verified that the statistical properties of the turbulent flow are also converged at the resolution we are using (see figure 3). 
\begin{figure}
\centering
\includegraphics[width = \columnwidth]{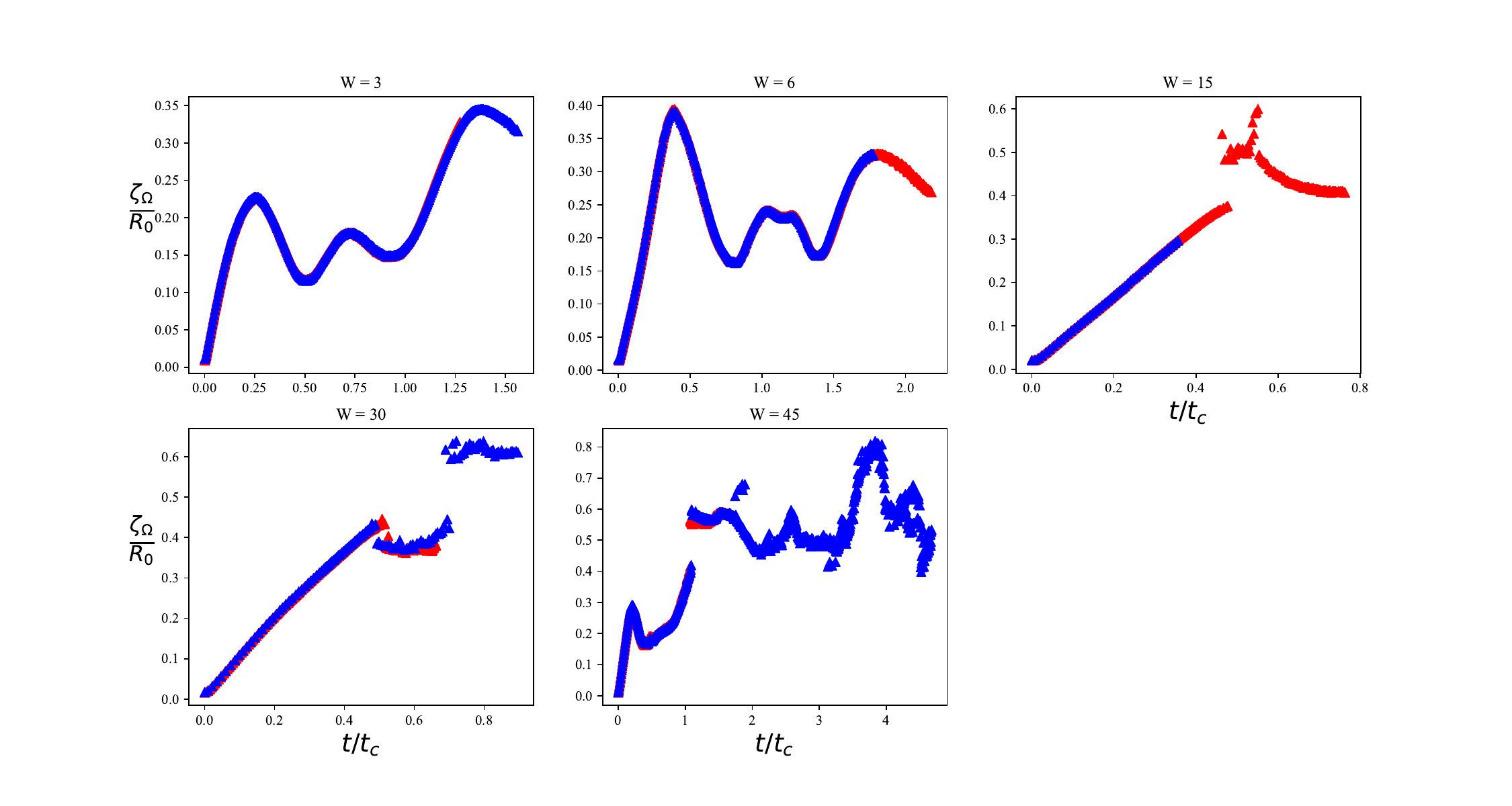}
\caption{Convergence test on $\zeta_\Omega$ as a function of dimensionless time $t/t_c$, for We=3,6,15,30,45. Level $N=9$~(blue) and level $N=10$ (red). Excellent convergence agreement is observed between the two numerical resolutions.}
\label{figconvergence}
\end{figure}

\begin{acknowledgments}
We are indebted to an anonymous reviewer for his comments on the presentation of the theoretical section in an earlier version of the paper. We thank St\'ephane Popinet for scientific discussion and development of the Basilisk package. We thank the three anonymous reviewers whose comments have helped improve the quality of the manuscript. This work was supported by the NSF CAREER award 1844932 to L.D., and the American Chemical Society Petroleum Research Fund Grant 59697-DNI9 to L.D. A.R. was supported by an International Fund grant from Princeton University to L.D. S.P. and A.R. were supported by the Labex ENS-ICFP.  We would like to acknowledge high-performance computing support from Cheyenne (doi:10.5065/D6RX99HX) provided by NCAR's Computational and Information Systems Laboratory, sponsored by the National Science Foundation. Computations were also performed on the Princeton supercomputer Tiger2, as well as on Stampede, through XSEDE allocations to L.D. and W.M., XSEDE is an NSF funded program 1548562. 
\end{acknowledgments}

Declaration of Interests. The authors report no conflict of interest.

\bibliographystyle{jfm}

\bibliography{biblio_bubbles}

\end{document}